%% file: tmNDMasses_prd_final.tex
\documentclass[aps,prd,preprintnumbers,showpacs,nofootinbib,floatfix]{revtex4}
\usepackage{amsmath,amssymb,amsbsy,mathbbol}
\usepackage{graphicx,subfigure,psfrag}

\newcommand{\transpose}{\mbox{${}^{\text{T}}$}}

\begin{document}

\include{def}

\preprint{UW-PT 05-07, UW-NT 05-01}

\title{Nucleon and Delta masses in twisted mass chiral perturbation
  theory} 

\author{Andr\'e Walker-Loud}
\email[]{walkloud@u.washington.edu}

\author{Jackson M. S. Wu}
\email[]{jmw@phys.washington.edu}
\affiliation{Department of Physics, University of Washington,
  Seattle, WA 98195-1560, USA}

\date{\today}

\begin{abstract}
We calculate the masses of the nucleons and deltas in twisted mass
heavy baryon chiral perturbation theory. We work to quadratic order in
a power counting scheme in which we treat the lattice spacing, $a$,
and the quark masses, $m_q$, to be of the same order. We give
expressions for the mass and the mass splitting of the nucleons and
deltas both in and away from the isospin limit. We give an argument
using the chiral Lagrangian treatment that, in the strong isospin
limit, the nucleons remain degenerate and the delta multiplet breaks
into two degenerate pairs to all orders in chiral perturbation
theory. We show that the mass splitting between the degenerate pairs
of the deltas first appears at quadratic order in the lattice
spacing. We discuss the subtleties in the effective chiral theory that
arise from the inclusion of isospin breaking.       
\end{abstract}

\pacs{12.38.Gc, 11.15.Ha, 12.39.Fe, 11.30.Rd}

\maketitle

%
%
%
%
%
%
%
%
%
%
%
%
\section{\label{sec:intro} Introduction}

Twisted mass lattice QCD (tmLQCD)~\cite{FetalLatt99,Fetal01} is an
alternative regularization for lattice QCD that has recently received
considerable attention.\footnote{For a recent review see
  Ref.~\cite{FrezLatt04}.} 
It has the potential to match the attractive features of improved
staggered fermions (efficient simulations~\cite{Kenn04}, absence of
``exceptional configurations''~\cite{FetalLatt99}, $O(a)$ improvement
at maximal twist~\cite{FR03}, operator mixing as in the
continuum~\cite{Fetal01,Pena04,FR04}) while not sharing the
disadvantage of needing to take roots of the determinant to remove
unwanted degrees of freedom. Thus tmLQCD offers a promising and
interesting new way to probe the properties and interactions of hadrons
non-perturbatively from first principles.

Due to the limitations in computational capabilities, the quark masses,
$m_q$, used in current simulations are still unphysically large. Thus
extrapolations in $m_q$ are necessary if physical predications are to
be made from lattice calculations. This can be done in a systematic
and model independent way through the use of chiral perturbation
theory ($\chi$PT). Since $\chi$PT is derived in the continuum, it can
be employed only after the continuum limit has been taken, where the
lattice spacing, $a$, is taken to zero. However, when close to the
continuum, it can be extended to lattice QCD at non-zero $a$, where
discretization errors arising from the finite lattice spacing are
systematically included in a joint expansion in $a$ and $m_q$. For
tmLQCD with mass-degenerate quarks, the resulting ``twisted mass
chiral perturbation theory'' (tm$\chi$PT) has been formulated
previously~\cite{MSch04,Scor04,SW04}, building on earlier work for the
untwisted Wilson theory~\cite{SS98,RS02,BRS03}.  

So far, tm$\chi$PT has only been applied to the mesonic (pionic)
sector. There have been studies on pion masses and decay constants
for $m_q\gg a\L_{\rm QCD}^2$~\cite{MSch04}, the phase structure of
tmLQCD for $m_q\sim a^2\L_{\rm
  QCD}^3$~\cite{Mun04,MunLatt04,Scor04,SW04,SWLatt04,AB04}, and
quantities involving pions that do not involve final state
interactions up to next-to-leading order (NLO) in the power counting
scheme where $m_q\sim a\L_{\rm QCD}^2$~\cite{SW04nlo}. However, as
pointed out in Ref.~\cite{SW04nlo}, many of the pionic quantities
considered are difficult to calculate in numerical simulations because
they involve quark-disconnected diagrams. This motivated us to extend
tm$\chi$PT to the baryon sector, which has heretofore not been done,
enabling us to study analytically baryonic quantities that do not
involve quark-disconnected diagrams. Numerical studies of the baryons in
tmLQCD are already underway, and the first results from quenched
simulations studying the nucleon and delta spectra have been obtained
recently in Ref.~\cite{ARLW05}. 

In the baryon sector, the extension of $\chi$PT to the lattice at
finite lattice spacing to $\c{O}(a)$~\cite{BeaSav03}, and to 
$\mathcal{O}(a^2)$~\cite{Tib05}, has been done for a theory with
untwisted Wilson fermions. We extend that work here to include the
effects of ``twisting'', i.e. our starting underlying lattice theory
is now tmLQCD. Specifically, we study the parity and flavor breaking
effects due to twisting in the masses and mass splittings of nucleons
and deltas in an $SU(2)$ chiral effective theory. The mass splittings
are of particular interest to us as they allow one to quantify the
size of the parity-flavor breaking effects in tmLQCD; furthermore,
they present less difficulties to numerical simulations than their 
counterparts in the mesonic sector, which involve quark-disconnected
diagrams.   

We consider here tmLQCD with mass non-degenerate quarks~\cite{FRLatt04},
which includes an additional parameter, the mass splitting,
$\epsilon_q$. This allows us to consider the theory both in and away
from the strong isospin limit. With simulations in the near future
most likely able to access the region where 
$m_q \sim a \Lambda_{\rm QCD}^2$, the power counting scheme we will
adopt is     
\begin{equation}
1 \gg \varepsilon^2 \sim a \Lambda_{\rm QCD} 
\sim \frac{m_q}{\Lambda_{\rm QCD}} 
\sim \frac{\epsilon_q}{\Lambda_{\rm QCD}}
\end{equation} 
with $\varepsilon^2$ denoting the small dimensionless expansion
parameters. In the following, we will work to $\c{O}(\varepsilon^4)$
in this power counting. 

The remainder of this article is organized as follows. In
Sec.~\ref{sec:EContL} and~\ref{sec:mesons}, we briefly review the 
definition of tmLQCD with mass non-degenerate quarks, and we show how
the mass splitting can be included in the Symanzik Lagrangian and the
$\c{O}(a)$ meson chiral Lagrangian. Higher order corrections from the
meson Lagrangian are not needed for the baryon observables to the order
we work. In Sec.~\ref{sec:baryons}, we extend the heavy baryon
$\chi$PT (HB$\chi$PT) to include the twisting effects to
$\c{O}(\varepsilon^4)$. In Sec.~\ref{sec:massiso} we present
the nucleon and delta masses in tm$\chi$PT in the strong isospin
limit, including lattice discretization errors and the flavor and
parity breaking induced by the twisted mass term.  
In Sec.~\ref{sec:mass} we extend the calculation to include
isospin breaking effects and discuss the subtleties that arise. We
conclude in Sec.~\ref{sec:conc}.

%
%
%
%
%
%
%
%
%
%
%
%
\section{\label{sec:tmChPT} Mass non-degenerate twisted mass chiral
  perturbation theory}

In this section, we work out the extension of the baryon chiral
Lagrangian to $\mathcal{O}(a^2)$ given in Ref.~\cite{Tib05} in
tmLQCD. We start by briefly outlining the construction of the Symanzik
Lagrangian in the mass non-degenerate case, which follows the same
procedures as those in the mass-degenerate theory with minimal
modifications. 

\subsection{\label{sec:EContL} The effective continuum quark level
  Lagrangian}

The fermionic part of the Euclidean lattice action of tmLQCD with two
mass non-degenerate quarks is
\begin{align} \label{E:ActionTw}
S^L_F =& \; \sum_{x} \bar{\psi_l}(x) \Big[
 \frac{1}{2} \sum_{\mu} \gamma_\mu (\nabla^\star_\mu + \nabla_\mu)
-\frac{r}{2} \sum_{\mu} \nabla^\star_\mu \nabla_\mu 
+ m_0 + i \gamma_5 \tau_1 \mu_0 - \tau_3 \epsilon_0
\Big] \psi_l(x),
\end{align}
where we have written the action given in Ref.~\cite{FRLatt04} for a general
twist angle (not necessarily maximal), and in the so-called ``twisted
basis''~\cite{FR03}. The quark (flavor) doublets $\psi_l$ and $\bar\psi_l$
are the dimensionless bare lattice fields (with ``$l$" standing for
lattice and not indicating left-handed), and $\nabla_\mu$ and 
$\nabla^\star_\mu$ are the usual covariant forward and backward
dimensionless lattice derivatives, respectively. The matrices $\tau_i$
are the usual Pauli matrices acting in the flavor space, with $\tau_3$
the diagonal matrix. The bare normal mass, $m_0$, the bare twisted
mass, $\mu_0$, and the bare mass splitting $\epsilon_0$, are all
dimensionless parameters; an implicit identity matrix in flavor space
multiplies the bare mass parameter $m_0$. The notation here is that
both $m_0$ and $\epsilon_0$ are positive such that the upper component
of the quark field is the lighter member of the flavor doublet with a
positive bare mass. 

Note that in the mass-degenerate case, twisting can be done using any
of the $\tau_i$, the choice of $\tau_3$ is merely for
convenience. Given the identity
\begin{equation}
\exp(-i\frac{\pi}{4}\tau_k)\,\tau_a\,\exp(i\frac{\pi}{4}\tau_k) 
= \epsilon_{kab} \tau_b \,,
\end{equation}
one can always rotate from a basis where the twist is implemented by
$\tau_a$, $a = 1,\,2$, to a basis where it is implemented by $\tau_3$
using the vector transformation 
\begin{equation}
\bar{\psi} \rightarrow \bar{\psi} \exp(i\theta\tau_k) \,, \qquad
\psi \rightarrow \exp(-i\theta\tau_k) \psi \,, \qquad\qquad
k = 1,\,2,\,3 \,, \qquad \theta = \pm \frac{\pi}{4} \,,
\end{equation}
where the appropriate sign for $\theta$ is determined by the index
$a$. However, with $\tau_3$ used here to split the quark doublet so
that the mass term is real and flavor-diagonal, it can not be used
again for twisting if the fermionic determinant is to remain
real.\footnote{%
One way to see this is to note that the mass terms 
$m_0 + i\gamma_5\tau_3\mu_0 - \tau_3\epsilon_0$ can be written as
$(x_0 - \tau_3 y_0)\exp(i\alpha\gamma_5)\exp(i\beta\gamma_5\tau_3)$,
where $x_0/y_0 = \tan\beta / \tan\alpha$. Thus, the twisted mass term
can be transformed away leaving just the normal mass term and the mass
splitting term. However, since this involves an $U(1)$ axial
transformation which is anomalous, 
an $i\alpha F\widetilde{F}$ term is introduced into the action which
we see now is complex (because of the factor of $i$). Thus, since the
gauge action is real, this means that the fermionic action (before the
transformation) must be complex, and so the fermionic determinant
obtained from it must also be complex. This also implies that a theory,
where both the twist and the mass
splitting are implemented by $\tau_3$, is $\alpha$-dependent.}      

Following the program of Symanzik~\cite{Sym83}, and the same
enumeration procedure detailed in Ref.~\cite{SW04}, one can obtain the
effective continuum Lagrangian at the quark level for mass
non-degenerate quarks that describes the long distance physics of the
underlying lattice theory. Its form is constrained by the symmetries
of the lattice theory. To $\mathcal{O}(\varepsilon^4)$ in our power
counting, in which we treat $a\Lambda_{\rm QCD}^2\sim m_q\sim\epsilon_q$,
we find that the Pauli term is again the only dimension five symmetry
breaking operator just as in the mass-degenerate case~\cite{SW04}
(the details of this argument are provided in Appendix~\ref{sec:appD5SB}),  
\begin{align} \label{E:CLeff}
\mathcal{L}_{\rm eff} &= \mathcal{L}_g + \bar{\psi} 
(D \!\!\!\!/ + m + i \gamma_5 \tau_1 \mu - \epsilon_q \tau_3) \psi 
+ b_1 a \bar{\psi}\, i \sigma_{\mu\nu} F_{\mu\nu}\, \psi 
+ O(a^2) \,,
\end{align}
where $\mathcal{L}_g$ is the continuum gluon Lagrangian, $m$ is the physical
quark mass, defined in the usual way by 
\begin{equation} 
m = Z_m(m_0 - \widetilde m_c)/a \,,
\end{equation}
$\mu$ is the physical twisted mass 
\begin{equation}
\mu = Z_\mu \mu_0/a = Z_P^{-1} \mu_0/a \,,
\end{equation}
and $\epsilon_q$ is the physical mass splitting
\begin{equation}
\epsilon_q = Z_\epsilon \epsilon_0/a = Z_S^{-1} \epsilon_0/a \,. 
\end{equation}
The factors $Z_P$ and $Z_S$ are matching factors for the non-singlet
pseudoscalar and scalar densities respectively. Note that the
lattice symmetries forbid additive renormalization to both $\mu_0$ and
$\epsilon_0$~\cite{FRLatt04}. The quantity $\tilde m_c$ is the
critical mass, aside from an $O(a)$ shift (see Ref.~\cite{SW04nlo} and
discussion below). 

Anticipating the fact that the mesons contribute to the baryon masses
only through loops, and so will be of $\c{O}(\varepsilon^3)$ or
higher, we only need to have a meson chiral Lagrangian to $\c{O}(a)$
for the order we work; $\c{L}_{\rm eff}$ as given in
Eq.~\eqref{E:CLeff} is sufficient for its construction. 
To build the effective chiral Lagrangian for baryons to
$\c{O}(\varepsilon^4)$ on the other hand, terms of $\c{O}(a^2)$ in
Eq.~\eqref{E:CLeff} are of the appropriate size to be
included. However, except for the operator which breaks $\c{O}(4)$
rotation symmetry, $a^2\bar{\psi}\gamma_\mu D_\mu D_\mu D_\mu\psi$,  
the $\c{O}(a^2)$ operators do not break the continuum symmetries in
a manner different than the terms explicitly shown in
Eq.~\eqref{E:CLeff}, and thus their explicit form is not needed. The
$\c{O}(4)$ breaking term will lead to operators in the baryon chiral
Lagrangian at the order we work. However, it is invariant under
twisting and thus contributes as those in the untwisted
theory~\cite{Tib05}.  

%
%
%
%
%
%
%
%
%
%
%
%
\subsection{\label{sec:mesons} The $SU(2)$ meson sector}

The low energy dynamics of the theory are described by a generalized
chiral Lagrangian found by matching from the continuum effective 
Lagrangian (\ref{E:CLeff}). As usual, the chiral Lagrangian is built
from the $SU(2)$ matrix-valued field $\Sigma$, which transforms under
the chiral group $SU(2)_L \times SU(2)_R$ as 
\begin{equation}
\Sigma \rightarrow L \Sigma R^{\dagger} \,, \qquad 
L \in SU(2)_L \,, \; R \in SU(2)_R \,. 
\end{equation}
The vacuum expectation value, $\Sigma_0 = \langle \Sigma \rangle$,
breaks the chiral symmetry spontaneously down to an $SU(2)$
subgroup. The fluctuations around $\Sigma_0$ correspond to the
pseudo-Goldstone bosons (pions).

From a standard spurion analysis, the chiral Lagrangian at
$\mathcal{O}(\varepsilon^2)$ is (in Euclidean space\footnote{We will
  work in Euclidean space throughout this article.}) 
\begin{equation} \label{E:ChiLeffLO}
\mathcal{L}_\chi = 
 \frac{f^2}{8} 
 \mathrm{Tr}(\partial_\mu \Sigma \partial_\mu \Sigma^\dagger)
-\frac{f^2}{8} 
 \mathrm{Tr}(\chi^{\dagger} \Sigma + \Sigma^\dagger\chi) 
-\frac{f^2}{8} 
 \mathrm{Tr}(\hat{A}^{\dagger} \Sigma + \Sigma^\dagger\hat{A}) \,,
\end{equation}
where $f$ is the decay constant (normalized so that $f_\pi = 132$
MeV). The quantities $\chi$ and $\hat{A}$ are spurions for the quark
masses and discretization errors respectively. At the end of the
analysis they are set to the constant values 
\begin{equation}
\chi \longrightarrow 2 B_0 (m + i\mu \tau_1 - \epsilon_q \tau_3) 
\equiv \hat{m} + i\hat{\mu} \tau_1 - \hat{\epsilon}_q \tau_3
\,, \qquad\qquad 
\hat A \longrightarrow 2 W_0 a \equiv \hat{a} \,,
\end{equation}
where $B_0\sim\c{O}(\L_{\rm QCD})$ and $W_0\sim\c{O}(\L_{\rm QCD}^3)$ 
are unknown dimensionfull constants, and we have defined the quantities
$\hat{m}$, $\hat{\mu}$ and $\hat{a}$.   

As explained in Ref.~\cite{SW04nlo}, since the Pauli term transforms
exactly as the quark mass term, they can be combined by using
the shifted spurion
\begin{equation}
\chi' \equiv \chi + \hat{A} \,,
\end{equation}
leaving the $\mathcal{O}(\varepsilon^2)$ chiral Lagrangian unchanged
from its continuum form. This corresponds at the quark level to a
redefinition of the untwisted component of the quark mass from $m$ to
\begin{equation} \label{E:m'def}
m' \equiv m + a W_0/B_0 \,.
\end{equation}
This shift corresponds to an $O(a)$ correction to the critical mass,
so that it becomes 
\begin{equation}
m_c = Z_m \tilde m_c / a - a W_0/B_0 \,.
\end{equation}

Since the $\mathcal{O}(\varepsilon^2)$ Lagrangian takes the continuum
form, and the mass splitting term does not contribute at this order,
the vacuum expectation value of $\Sigma$ at this order is that which
cancels out the twist in the shifted mass matrix, exactly as in the
mass-degenerate case:   
\begin{equation} \label{E:VacLO}
\langle0|\Sigma|0\rangle_{LO} \equiv \Sigma_{0} = 
\frac{\hat{m} + \hat{a} + i \hat{\mu} \tau_1}{M'}
\equiv \exp(i \omega_0 \tau_1) \,,
\end{equation}
where
\begin{equation} 
\label{E:M'def}
M' = \sqrt{(\hat{m}+\hat{a})^2 +\hat{\mu}^2} \,.
\end{equation}
Note that $M'$ is the leading order result for the pion mass-squared,
i.e. $m_\pi^2 = M'$ at $\mathcal{O}(\varepsilon^2)$. If we define the
physical quark mass by 
\begin{equation}
m_q = \sqrt{m'^2 + \mu^2} \,,
\end{equation}
then it follows from (\ref{E:VacLO}) that
\begin{equation}
\cos\omega_0 = m'/m_q \,,\qquad
\sin\omega_0 = \mu /m_q \,.
\end{equation}
Details of the non-perturbative determination of the twist angle and the
critical mass can be found in~\cite{SW04nlo}, and will not be repeated
here.

At $\mathcal{O}(\varepsilon^4)$, the mass non-degenerate chiral
Lagrangian for the pions retains the same form as that in the mass
degenerate case~\cite{SW04,SW04nlo}, because the mass splitting does
not induce any additional symmetry breaking operators in
$\mathcal{L}_{\rm eff}$. The $\mathcal{O}(\varepsilon^4)$ pion
Lagrangian contains the usual Gasser-Leutwyler operators of 
$\mathcal{O}(\mathsf{m}^2,\,\mathsf{m}p^2,\,p^4)$, where $\mathsf{m}$ 
is a generic mass parameter that can be $m,\,\mu$, or $\epsilon_q$, as
well as terms of $\mathcal{O}(a\mathsf{m},\,a p^2,\,a^2)$ associated
with the discretization errors. Now as we stated earlier, since the
pions will enter only through loops in typical calculations of baryon
observables, keeping the pion masses to $\mathcal{O}(\varepsilon^4)$
will lead to corrections of $\mathcal{O}(\varepsilon^5)$, which is
beyond the order we work. As our concern is not in the meson sector,
the $\mathcal{O}(\varepsilon^2)$ pion Lagrangian~(\ref{E:ChiLeffLO})
is thus sufficient for our purpose in this work.    

%
%
%
%
%
%
%
%
%
%
%
%
\subsection{\label{sec:baryons} The $SU(2)$ baryon sector}

With the effective continuum theory and the relevant part of the
effective chiral theory describing the pions in hand, we now include
the lowest lying spin-$\frac{1}{2}$ and spin-$\frac{3}{2}$ baryons
into tm$\chi$PT by using HB$\chi$PT~\cite{JM91t,JM91,Jen92}, which we 
will refer as the twisted mass HB$\chi$PT (tmHB$\chi$PT). In $SU(2)$
the spin-$\frac{1}{2}$ nucleons are described by a doublet 
\begin{equation} \label{E:Nfield}
N = 
\begin{pmatrix}
p \\
n
\end{pmatrix}\,,
\end{equation}
and the delta resonances form a flavor quartet. As they are
spin-$\frac{3}{2}$, they are described by a Rarita-Schwinger
field, $T_\mu^{ijk}$, which is totally symmetric in flavor and
satisfies $\gamma_\mu T_\mu = 0$. The delta fields are normalized such
that 
\begin{equation} \label{E:Tfield}
T^{111} = \Delta^{++} \,, \quad 
T^{112} = \frac{1}{\sqrt{3}}\Delta^{+} \,, \quad
T^{122} = \frac{1}{\sqrt{3}}\Delta^{0} \,, \quad 
T^{222} = \Delta^{-} \,.
\end{equation}
The free Lagrangian for the nucleons and deltas to
$\mathcal{O}(\varepsilon^2)$ consistent with the symmetries of the
lattice theory is (in Euclidean space) 
\begin{align} \label{E:BLLOtw}
\c{L}_\chi =
&\,\ol{N} i v \cdot D N 
-2\,\a_M\,\ol{N} \c{M}_+^{tw} N
-2\,\s_M\,\ol{N} N \,{\rm tr}(\c{M}_+^{tw})
-2\,\s_W\,\ol{N} N \,{\rm tr}(\c{W}_+) \notag \\
& + (\ol{T}_\mu i v \cdot D\,T_\mu) 
  + \Delta\,(\overline{T}_\mu T_\mu) 
  +2\,\gamma_M\,(\overline{T}_\mu \mathcal{M}_+^{tw} T_\mu)
  -2\,\ol{\sigma}_M\,(\ol{T}_\mu T_\mu)\,{\rm tr}(\mathcal{M}_+^{tw})  
  -2\,\ol{\sigma}_W\,(\ol{T}_\mu T_\mu)\,{\rm tr}(\mathcal{W}_+) \,,
\end{align}
where the trace is taken in flavor space, and the notation,
(\,$\ldots$\,), denotes contractions of the flavor (tensor) indices as
defined in e.g. Ref.~\cite{LS96}. The ``twisted mass'' spurion field
is defined by    
\begin{equation}
\mathcal{M}_{\pm}^{tw} = \frac{1}{2} \left[ 
\xi^\dagger m_Q^{tw} \xi^\dagger \pm \xi (m_Q^{tw})^\dagger \xi
\right] \,, \qquad\qquad
m_Q^{tw} = \frac{\chi'}{2B_0} \,,
\end{equation}
with $m_Q^{tw}$ being the ``twisted'' mass spurion for the
baryons. The ``Wilson'' (discretization) spurion field is defined by
\begin{equation}
\mathcal{W}_{\pm} = \frac{1}{2} \left( 
\xi^\dagger w_Q \xi^\dagger \pm \xi w_Q^\dagger \xi 
\right) \,, \qquad\qquad
w_Q = \frac{\Lambda_{\rm QCD}^2}{2W_0}\hat{A} \,,
\end{equation}
with $w_Q$ being the Wilson spurion for the baryons. Note that we have
made simplifications using the properties of $SU(2)$ matrices when
writing down Eq.~\eqref{E:BLLOtw}. When setting the spurions to their
constant values, $\mathcal{W}_+$ is proportional to the identity
matrix in flavor space. Thus the operators $\ol{N}\c{W}_+ N$ and 
$(\ol{T}_\mu\c{W}_+ T_\mu)$, although allowed under the symmetries of
tmLQCD, are not independent operators with respect to
$\ol{N}N\,{\rm tr}(\c{W}_+)$ and $(\ol{T}_\mu T_\mu)\,{\rm tr}(\c{W}_+)$
respectively. This is also true of the nucleon and delta operators
involving $\c{M}_+^{tw}$ in the isospin limit (but not away from it). The
independent operators we choose to write down are those with the
simplest flavor contractions, and this will be the case henceforth
whenever we make simplifications using the properties of $SU(2)$.   

In Eq.~(\ref{E:BLLOtw}), the four-vector, $v_\mu$, is the heavy baryon
four-velocity, and our conventional here is that $v \cdot v = 1$. The
parameter, $\D$, is the mass splitting between the nucleons and deltas
which is independent of the quark masses (often referred to as the
nucleon-delta mass splitting in the chiral limit), and we treat 
$\Delta \sim m_\pi \sim \varepsilon^2$
following~\cite{JM91t,JM91,Jen92,HHK98}. The chiral covariant 
derivative, $D_\mu$, acts on the nucleon and delta fields as  
\begin{align}
(D_\mu N)_i &= 
\partial_\mu N_i + (\mathcal{V}_\mu)_i^{\;\;j} N_j \,, 
\notag \\ 
(D_\mu T^\nu)_{ijk} &= 
\partial_\mu T^\nu_{ijk}
+(\mathcal{V}_\mu)_i^{\;\;i'} T^\nu_{i'jk}
+(\mathcal{V}_\mu)_j^{\;\;j'} T^\nu_{ij'k}
+(\mathcal{V}_\mu)_k^{\;\;k'} T^\nu_{ijk'} \,.
\end{align}
The vector and axial-vector fields are defined by
\begin{equation}
\mathcal{V}_\mu = \frac{1}{2}\left(
\xi \partial_\mu \xi^\dagger + \xi^\dagger\partial_\mu \xi
\right) \,, \qquad
\mathcal{A}_\mu = \frac{i}{2}\left(
\xi \partial_\mu \xi^\dagger - \xi^\dagger\partial_\mu \xi
\right) \,, \qquad
\xi^2 = \Sigma \,.
\end{equation}
The dimensionless low energy constants (LECs), $\alpha_M$, $\sigma_M$,
$\gamma_M$, and $\overline{\sigma}_M$ have the same numerical values
as in the usual untwisted two-flavor HB$\chi$PT. 

As was noted in Ref.~\cite{SW04nlo}, the shifting from $\chi$ to
$\chi' = \chi + \hat{A}$, which corresponds to the shift of the
physical mass $m$ to $m'$ at the quark level does not, in general, 
remove the discretization ($\hat{A}$) term, and this is seen
explicitly here with the presence of the discretization terms. 

The Lagrangian describing the interactions of the nucleons and deltas
with the pions is  
\begin{equation} \label{E:BLTN}
\c{L}_\chi = 
2\,g_A \,\overline{N} S \cdot A \, N - 
2\,g_{\Delta\Delta}\,\ol{T}_\mu S\cdot\c{A}\,T_\mu +  
g_{\Delta N}\,\left[\ol{T}^{kji}_\mu\c{A}_i^{\mu,i^\prime}
                    \epsilon_{j i^\prime} N_k + h.c.\right] \,,
\qquad i,\,j,\,k = 1,\,2,\,3 \,.
\end{equation}
The tensor $\epsilon_{ij}$ is the rank-2 analogue of the totally
antisymmetric tensor $\epsilon_{ijk}$. The vector $S_\mu$ is the
covariant spin operator~\cite{JM91t,JM91}. The LECs in (\ref{E:BLTN})
are the same as those in the untwisted two-flavor HB$\chi$PT. Note
that the $\c{O}(\varepsilon^2)$ free Lagrangian (\ref{E:BLLOtw}) and
interaction Lagrangian (\ref{E:BLTN}) are the same as those given in
Ref.~\cite{Tib05} when the twist is removed, i.e. when $\mu = 0$. With
non-vanishing twist, the mass operators carry a twisted component and
the vacuum is ``twisted'' from the identity to point in the direction
of the twist (the flavor $\tau_1$-direction here)~\cite{SW04nlo}.   

\bigskip

Following Ref.~\cite{SW04nlo}, we expand $\Sigma$ about its vacuum
expectation value, defining the physical pion fields and the physical
$\xi$ fields by  
\begin{equation} \label{E:LOexpand}
\Sigma = \c{T}\Sigma_{ph}\c{T} \,, \qquad
\xi = \c{T}\xi_{ph} V(\xi_{ph}) \,, \qquad
\c{T} = \exp(i\omega_0 \tau_1 / 2) \,, \qquad
\Sigma_{ph} =
\exp(i\sqrt{2}\;\boldsymbol{\pi}\cdot\boldsymbol{\tau}/f)
\,, \qquad
\c{T} \,, V \in SU(2) \,,
\end{equation}
Now, if we make the following chiral transformation (under which the
effective chiral Lagrangian is invariant)
\begin{align} \label{E:ChiTrans}
\Sigma \rightarrow L \Sigma R^\dagger & \,,\qquad
\xi \rightarrow L \xi V^\dagger \equiv V \xi R^\dagger \,, \qquad  
N^i \rightarrow V^{i j} N^j \,, \qquad 
T^{ijk}_\mu \rightarrow V^{ii'}V^{jj'}V^{kk'}T^{i'j'k'}_\mu \,,
\notag \\
&\chi' \rightarrow L \chi' R^\dagger \,, \qquad 
\hat{A} \rightarrow L \hat{A} R^\dagger \,,
\qquad L \,,\, R \,,\, V \in SU(2) \,,
\end{align}
using the particular $SU(2)$ matrices $L = R^\dagger = \c{T}^\dagger$,
we have in the transformed effective chrial Lagrangian
\begin{equation} \label{E:PhysFields}
\Sigma \rightarrow \Sigma_{ph} \,, \qquad 
\xi \rightarrow \xi_{ph} \,, \qquad 
\mathcal{A}_\mu \rightarrow 
\frac{i}{2}\left(\xi_{ph} \partial_\mu \xi_{ph}^\dagger -
            \xi_{ph}^\dagger \partial_\mu \xi_{ph}\right) 
\,, \qquad 
\mathcal{V}_\mu \rightarrow 
\frac{1}{2}\left(\xi_{ph} \partial_\mu \xi_{ph}^\dagger +
            \xi_{ph}^\dagger \partial_\mu \xi_{ph}\right) \,,
\end{equation}
and
\begin{align}
\mathcal{M}^{tw}_\pm &\rightarrow \mathcal{M}_\pm = \frac{1}{2} 
\left(\xi_{ph}^\dagger m_Q \xi_{ph}^\dagger \pm 
      \xi_{ph} m_Q^\dagger \xi_{ph} \right) \,, 
& m_Q^{tw} &\rightarrow m_Q = 
\c{T}^\dagger \frac{\chi'}{2B_0} \c{T}^\dagger \,, \notag \\
\mathcal{W}_\pm &\rightarrow \mathcal{W}^{tw}_\pm = \frac{1}{2} 
\left[\xi_{ph}^\dagger w_Q^{tw} \xi_{ph}^\dagger \pm 
      \xi_{ph} (w_Q^{tw})^\dagger \xi_{ph} \right] \,, 
& w_Q &\rightarrow w_Q^{tw} = 
\c{T}^\dagger
\left(\frac{\Lambda_{\rm QCD}^2}{2W_0}\hat{A}\right)
\c{T}^\dagger \,.
\end{align}
Note that since $L = R^\dagger = \c{T}^\dagger \in SU(2)$, and
$\xi = L^\dagger\xi_{ph}V \equiv V^\dagger\xi_{ph}R$,
\begin{equation}
\Sigma = L^\dagger\Sigma_{ph}R = 
 \xi^2 = (L^\dagger \xi_{ph} V)\cdot(V^\dagger \xi_{ph} R) 
       = L^\dagger\xi_{ph}^2 R
\,\Longrightarrow\,\xi_{ph}^2 = \Sigma_{ph} \,.
\end{equation}

We see that the $\xi$ field is now $\xi_{ph}$, the field associated
with the physical pions, and the twist is transferred from the twisted
mass ($\mathcal{M}^{tw}_\pm$) term to the ``twisted Wilson''
($\mathcal{W}^{tw}_\pm$) term, making the mass term in the HB$\chi$PT
now the same as that in the untwisted theory. The new mass spurion,
$m_Q$, and the ``twisted Wilson'' spurion, $w_Q^{tw}$, now take
constant values  
\begin{equation}
m_Q \longrightarrow m_q - \epsilon_q \tau_3 
\,, \qquad\qquad 
w_Q^{tw} \longrightarrow 
a\,\Lambda_{QCD}^2 \exp(-i\omega_0 \tau_1) \,.
\end{equation}
We will call this the ``physical pion basis'' since this is the basis
where the pions are physical~\cite{SW04nlo}, and we will work in this
basis from now on, unless otherwise specified.\footnote{As detailed in
  Ref.~\cite{SW04nlo}, the twist angle that one determines
  non-perturbatively in practice, call it $\omega$, will differ from
  $\omega_0$ by $O(a)$. This will give rise to a relative $O(a)$
  contribution to the pion terms. But since the pions come into baryon
  calculations only through loops, the correction will be of higher
  order than we work. Thus to the accuracy we work, we may use either
  $\omega$ or $\omega_0$.}  
A technical point we note here is that, in the isospin limit where
twisting can be implemented by any of the three Pauli matrices, say
$\tau_k$, the physical pion basis can be found following the same
recipe detailed above but with $\tau_1$ in $\c{T}$ replaced by
$\tau_k$ throughout. 

Rotating to the physical pion basis where the $\xi$ field is now the
physical $\xi_{ph}$ field in all field quantities, the form of the
interaction Lagrangian remains unchanged as given in (\ref{E:BLTN}),
while the $\c{O}(\varepsilon^2)$ free heavy baryon chiral
Lagrangian~(\ref{E:BLLOtw}) changes to   
\begin{align} \label{E:BLLOphy}
\c{L}_\chi = 
&\,\ol{N} i v \cdot D N 
-2\,\a_M\,\ol{N}\c{M}_+ N
-2\,\s_M\,\ol{N}N\,{\rm tr}(\c{M}_+)
-2\,\s_W\,\ol{N}N\,{\rm tr}(\c{W}^{tw}_+) \notag \\
& + (\ol{T}_\mu i v \cdot D\,T_\mu) 
  +\Delta\,(\ol{T}_\mu T_\mu) 
  +2\,\g_M\,(\ol{T}_\mu \c{M}_+ T_\mu)
  -2\,\ol{\s}_M\,(\ol{T}_\mu T_\mu)\,{\rm tr}(\c{M}_+)
  -2\,\ol{\s}_W\,(\ol{T}_\mu T_\mu)\,{\rm tr}(\c{W}^{tw}_+) \,.
\end{align}
Note that $\c{W}^{tw}_+$ is also proportional to the identity matrix
in flavor space when set to its constant value. Thus if we build
the free chiral Lagrangian directly in the physical pion basis, the
same simplifications due to $SU(2)$ we used in writing down
Eq.~(\ref{E:BLLOtw}) apply. Note that at this point, one can not yet tell
whether the nucleon ($N$) and the delta ($T_\mu$) fields are
physical. This has to be determined by the theory itself. We will
return to this point when calculating the nucleon and delta masses
below. 

\bigskip

At $\mathcal{O}(\varepsilon^4)$, there are contributions from
$\mathcal{O}(a \mathsf{m})$ and $\mathcal{O}(a^2)$ operators. The
enumeration of the operators is similar to that set out in
Ref.~\cite{Tib05}, except now the Wilson spurion field carries a
twisted component. The operators appearing in the
$\mathcal{O}(\varepsilon^4)$ chiral Lagrangian will involve two 
insertions off the following: $\mathcal{M}_\pm$,
$\mathcal{W}^{tw}_\pm$, and the axial current $\mathcal{A}_\mu$. Note
that since parity combined with flavor is conserved in tmLQCD, any one
insertion of $\mathcal{M}_-$ or $\mathcal{W}^{tw}_-$ must be
accompanied by another insertion of $\mathcal{M}_-$ or
$\mathcal{W}^{tw}_-$. Now operators with two insertions of 
$\mathcal{M}_+$ or $\mathcal{A}_\mu$, which contribute to baryon
masses at tree and one-loop level respectively, have the same form as
those in the untwisted theory (and so give the same
contribution). These have been written down in~\cite{TWL05b} and will
not be repeated here. Operators with an insertion of either a
combination of $v \cdot \c{A}$ and $\c{M}_-$ (which have the same form
as in the untwisted theory), or a combination of $v \cdot \mathcal{A}$
and $\mathcal{W}^{tw}_-$, will also not contribute to the baryon
masses at $\mathcal{O}(\varepsilon^4)$.    

At $\mathcal{O}(a \mathsf{m})$, there are two independent operators 
contributing to the masses in the nucleon sector
\begin{align} \label{E:Namq}
\mathcal{L}_\chi &= -\frac{1}{\Lambda_\chi} \bigg[
n_1^{WM_+}\,\overline{N}\mathcal{M}_+ N\,{\rm tr}(\mathcal{W}^{tw}_+) +
n_2^{WM_+}\,\overline{N}N
          \,{\rm tr} (\mathcal{M}_+) \,{\rm tr}(\mathcal{W}^{tw}_+)
\bigg] \,,
\end{align}
and two independent operators contributing to the masses in the delta
sector 
\begin{align} \label{E:Tamq}
\mathcal{L}_\chi &= \frac{1}{\Lambda_\chi} \bigg[
t_1^{WM_+}\,(\overline{T}_\mu \mathcal{M}_+ T_\mu)
          \,{\rm tr}(\mathcal{W}^{tw}_+) +
t_2^{WM_+}\,(\overline{T}_\mu T_\mu)
          \,{\rm tr}(\mathcal{W}^{tw}_+)\,{\rm tr}(\mathcal{M}_+)
\bigg] \,, 
\end{align}
where $\L_\chi\equiv4\pi f$.\footnote{We will use this definition as our
  convention, which differs from the more standard convention,
  $\L_\chi=2\sqrt{2}\pi f$, employed by other authors.}   
Note that there are no operators involving the commutator,
$[\mathcal{M}_+,\mathcal{W}^{tm}_+]$, because it is identically
zero. There are also operators involving $\c{M}_- \otimes\c{W}^{tw}_-$
at $\c{O}(a\mathsf{m})$, but these again do not contribute to the
baryon masses at the order we work. 

Note that the scale used in Eqs.~\eqref{E:Namq} and~\eqref{E:Tamq}
above to make the dimensions correct is not the QCD scale, 
$\L_{\rm QCD}$, but the $\chi$PT scale, $\L_\chi$. This follows from
the naive dimensional analysis (NDA) of Ref.~\cite{MG1984}, and from
the fact that these operators also function as counter terms for
divergences arising from the leading loop diagrams, which have
contributions proportional to $\c{O}(a\mathsf{m})$.  
We use the same analysis to set the dimensionfull scale in all the
following operators which contribute to the nucleon and delta masses
to the order we work. 

At $\mathcal{O}(a^2)$, there are operators that do not break the 
chiral symmetry arising from the bilinear operators and four-quark
operators (see e.g. Ref.~\cite{BRS03} for a complete listing) in 
the $\mathcal{O}(a^2)$ part of $\mathcal{L}_{\rm eff}$. These give
rise to the tmHB$\chi$PT operators
\begin{equation} \label{E:Na2chi}
\c{L}_\chi = a^2\L^3_{\rm QCD}
\bigg[-b\,\ol{N}N+t\,(\ol{T}_\mu T_\mu) \bigg] \,.
\end{equation}
There are also chiral symmetry preserving but $O(4)$ rotation symmetry
breaking operators which arise from the bilinear operator of 
the form $a^2\bar{\psi}\gamma_\mu D_\mu D_\mu D_\mu \psi$ in the
$\c{O}(a^2)$ part of $\mathcal{L}_{\rm eff}$. These give rise to the
tmHB$\chi$PT operators 
\begin{equation} \label{E:Na2O4}
\c{L}_\chi = a^2\L^3_{\rm QCD} \bigg[
-b_v\,\ol{N} v_\mu v_\mu v_\mu v_\mu N
+t_v\,(\ol{T}_\nu v_\mu v_\mu v_\mu v_\mu T_\nu)
+t_{\bar{v}}\,(\ol{T}_\mu v_\mu v_\mu T_\mu) \bigg] \,.
\end{equation}
Note that these chiral symmetry preserving operators are clearly not
affected by twisting (the $O(4)$ symmetry breaking operator at the
quark level from which they arise involve only derivatives with no
flavor structure, and $\{\gamma_\mu,\gamma_5\} = 0$), and so they have 
the same form and contribute to the baryon masses in the same way as
in the untwisted theory. The chiral symmetry breaking operators at
$\mathcal{O}(a^2)$ are those with two insertions of the Wilson spurion
fields. For the nucleons, there are two such independent operators     
\begin{equation} \label{E:Nasq}
\mathcal{L}_\chi = -\frac{1}{\L_{\rm QCD}} \bigg[
n_1^{W_+}\,\overline{N}N
         \,{\rm tr}(\mathcal{W}^{tw}_+)\,{\rm tr}(\mathcal{W}^{tw}_+) + 
n_1^{W_-}\,\overline{N}N 
         \,{\rm tr}(\mathcal{W}^{tw}_-\mathcal{W}^{tw}_-) 
\bigg] \,,
\end{equation}
and for the deltas, there are three such independent operators
\begin{equation} \label{E:Tasq}
\mathcal{L}_\chi = \frac{1}{\L_{\rm QCD}} \bigg[
t_1^{W_+}\,(\overline{T}_\mu T_\mu)
         \,{\rm tr}(\mathcal{W}^{tw}_+)\,{\rm tr}(\mathcal{W}^{tw}_+) +
t_1^{W_-}\,(\overline{T}_\mu T_\mu)
         \,{\rm tr}(\mathcal{W}^{tw}_- \mathcal{W}^{tw}_-) +
t_2^{W_-}\,\overline{T}^{kji}_\mu
           (\mathcal{W}^{tw}_-)^{ii'}(\mathcal{W}^{tw}_-)^{jj'}
           T^{i^\prime j^\prime k}_\mu
\bigg] \,.
\end{equation}

In the isospin limit where the mass splitting vanishes ($\epsilon_q
\rightarrow 0$), more simplifications occur in the
$\c{O}(\varepsilon^4)$ chiral Lagrangian. The nucleon operators with
coefficients $n_1^{WM_+}$ and $n_2^{WM_+}$ are the same up to a
numerical factor, and the same holds for delta operators with
coefficients $t_1^{WM_+}$, $t_2^{WM_+}$, $t_3^{WM_+}$, and for
operators with coefficient $t_1^{W_-}$ and $t_2^{W_-}$. 

Note that in the untwisted limit, the $\mathcal{O}(\varepsilon^4)$
chiral Lagrangian reduces to that given in Ref.~\cite{Tib05}. In
particular, with the twist set to zero, operators with two insertions of
$\mathcal{W}^{tw}_-$ will not contribute to the nucleon or the delta
mass until $\c{O}(a^2\mathsf{m}) \sim \c{O}(\varepsilon^6)$. But for  
non-vanishing twist, they contribute at $\mathcal{O}(a^2)$.

%
%
%
%
%
%
%
%
%
%
%
%
%
\section{\label{sec:massiso} Nucleon and delta masses in the isospin
  limit} 

There have been extensive studies of the nucleon masses, and to a
lesser extent, the delta masses in HB$\chi$PT and other variants of
$\chi$PT. A partial list of references of these studies
includes~\cite{Jen92,HHK98,WL05,TWL05a,TWL05b,Tib05,Tib05PQ,Bernard93,
  Lebed94a,Lebed94b,Banerjee95,Borasoy97,Becher99,Frink04,Lehnhart04}.
In this work, we are concerned with corrections to the masses of the
nucleons and the deltas due to the effect of the twisted mass
parameter. We will therefore only give expressions for the mass
corrections arising from the effects of lattice discretization and
twisting in tmLQCD. A calculation of the nucleon and delta masses in
the continuum in infinite volume to $\c{O}(m_q^2)$ can be found in
Ref.~\cite{TWL05b}. The mass corrections due to finite lattice spacing
to $\c{O}(a^2)$ in the untwisted theory with Wilson quarks can be
found in Ref.~\cite{Tib05}, and the leading finite volume
modifications to the nucleon mass can be found in Ref.~\cite{Beane04}.

In this section, we present the results of nucleon and delta masses
calculated in tmHB$\chi$PT, in the isospin limit, where the quark
doublet is mass-degenerate, and the twist is implemented by
$\tau_3$. As we discussed in Sec.~\ref{sec:EContL}, in the isospin
limit, the content of tmLQCD is the same regardless of which Pauli
matrix is used to implement the twist -- the action for one choice is
related to another by a flavor-vector rotation. This must also hold true of
the effective chiral theory that arises from tmLQCD. Indeed, the heavy
baryon Lagrangian constructed in Sec.~\ref{sec:baryons} with
$\tau_1$-twisting can be rotated into that with $\tau_3$-twisting by 
making a vector transformation, which is given by Eq.~(\ref{E:ChiTrans}) 
but with  $L = R = V = \exp(i\frac{\pi}{4}\tau_2)$. 

\subsection{\label{sec:NMassIso} Nucleon Masses in the Isospin Limit}

In the continuum, the mass of the nucleons in infinite volume
HB$\chi$PT with two flavor-degenerate quarks are organized as an
expansion in powers of the quark mass, which can be written
as\footnote{Here we use a different convention from some of the more
  recent nucleon mass calculations.}
\begin{equation} \label{eq:Nmassexp}
M_{N_i} = M_0\left(\D,\L_R\right) 
        + M_{N_i}^{(1)}\left(\D,\L_R\right) 
        + M_{N_i}^{(3/2)}\left(\D,\L_R\right) 
        + M_{N_i}^{(2)}\left(\D,\L_R\right) + \ldots
\end{equation}
where $N_i$ stands for either the proton $(i=p)$ or the neutron
$(i=n)$, and $M_{N_i}^{(n)}$ is the contribution to the $i^{th}$
nucleon of $\c{O}(m_q^n)$ calculated in the continuum and infinite
volume two-flavor $\chi$PT in the isospin limit~\cite{TWL05b}. The
quantity, $M_0\left(\D,\L_R\right)$, is the renormalized nucleon 
mass in the chiral limit; it is independent of $m_q$ and $N_i$, but
depends the renormalization scale, $\L_R$, and on $\D$, the renormalized 
mass splitting between the nucleons and deltas which is independent of
the quark masses. Here, we are interested in the corrections to this
formula due to the effects of lattice discretization and twisting
arising from tmLQCD. We denote these lattice corrections to the
nucleon mass at $\c{O}(\varepsilon^{2n}) \sim \c{O}(m_q^n) \sim
\c{O}(a^n)\,$ (factors of $\L_{\rm QCD}$ needed to make the dimensions
correct are implicit here) as $\d M_{N_i}^{(n)}$, and the nucleon mass
in tmHB$\chi$PT is now written as 
\begin{equation}
M_{N_i}^{tm} = M_0 + (M_{N_i}^{(1)} + \delta M_{N_i}^{(1)}) + \ldots
\end{equation}
Throughout this work, we use dimensional regularization with a
modified minimal subtraction scheme where we consistently subtract off
terms proportional to 
\begin{equation}
\frac{1}{\varepsilon} - \g_E + 1 + \log 4 \pi \,.
\end{equation}

The leading correction in tm\CPT\ comes in at tree
level, arising from the twisted Wilson nucleon operator in the
free heavy baryon Lagrangian~(\ref{E:BLLOphy}). It reads
\begin{equation} \label{E:NBLO}
\d M_{N_i}^{(1)}(\w) = -4\,\s_W\,a\,\L_{\rm QCD}^2\cos(\w) \,,
\end{equation}
where to the accuracy we work, $\omega$ can either be $\omega_0$ or
the twist angle non-perturbatively determined. Note that this
correction is the same for both the proton and the neutron. At leading
order, the nucleon mass is automatically $\c{O}(a)$ improved, as 
$\d M_{N_i}^{(1)}$ vanishes at maximal twist, $\omega = \pi/2$. At
zero twist, $\omega = 0$, it reduces to that in the untwisted
theory~\cite{Tib05,BeaSav03}.  

\begin{figure}[t]
\centering
\includegraphics[width=0.2\textwidth]{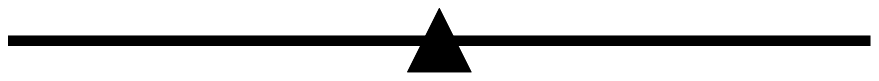}
\hspace{0.05\textwidth}
\includegraphics[width=0.2\textwidth]{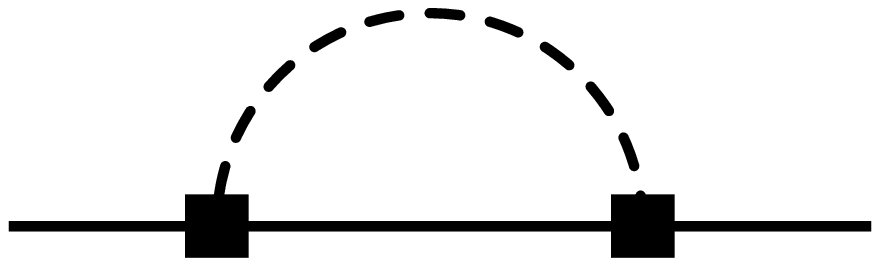}
\hspace{0.05\textwidth}
\includegraphics[width=0.2\textwidth]{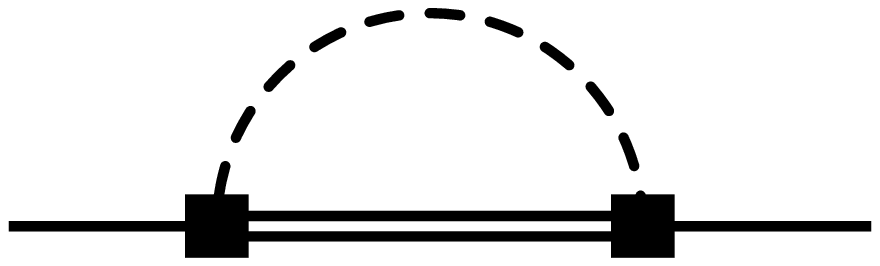}
\caption{Diagrams depicting LO and NLO mass contributions to the
  nucleons in tmHB$\chi$PT in the physical pion basis. The solid,
  double solid and dashed lines denote nucleons, deltas and pions
  respectively. The solid triangle represents an insertion of the
  twisted Wilson operator as given in Eq.~\eqref{E:BLLOphy}. The solid
  squares denote the couplings of the baryons to the axial current
  whose form is given in Eq.~\eqref{E:BLTN}.}  
\label{fig:LOandNLO}
\end{figure}

The next contribution to the nucleon mass comes from the leading pion
loop diagrams shown in Fig.~\ref{fig:LOandNLO}. However, at the order
we work, the form of the $\c{O}(\mathsf{m}^{3/2})$ nucleon mass
contribution is unchanged from the continuum. For completeness we give
its full expression here 
\begin{equation} \label{eq:MBNLO}
M^{(3/2)}_{N_i} = 
-\frac{3}{16\pi f^2} g_A^2  m_\pi^3
-\frac{8g_{\D N}^2}{3(4 \pi f)^2}\c{F}(m_\pi,\D,\L_R) \,,
\end{equation}
where $m_\pi^2 = M' = 2B_0\,m_q$, is the physical pion mass-squared
given in Eq.~\eqref{E:M'def}, and the function $\c{F}$ is given by 
\begin{equation} \label{E:F}
\c{F}(m,\d,\L_R) = 
(m^2 - \d^2)\left[\sqrt{\d^2 - m^2}\log\left(
\frac{\d - \sqrt{\d^2 - m^2 + i \varepsilon}}
     {\d + \sqrt{\d^2 - m^2 + i \varepsilon}} \right)- 
\d\log\left(\frac{m^2}{\L_R^2} \right)\right] - 
\frac{\d}{2}\,m^2\log\left(\frac{m^2}{\L_R^2}\right).
\end{equation}

\begin{figure}[t] 
\centering
\includegraphics[width=0.2\textwidth]{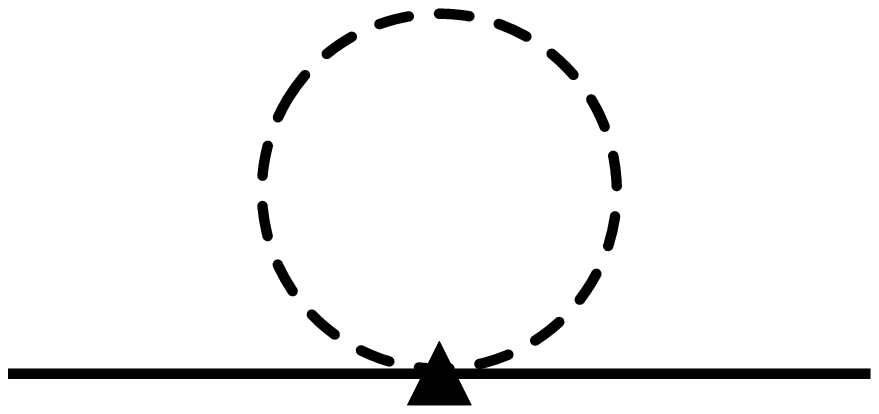}
\hspace{0.05\textwidth}
\includegraphics[width=0.2\textwidth]{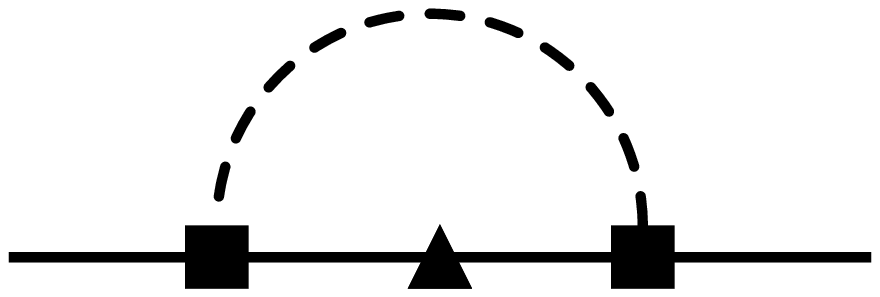}
\hspace{0.05\textwidth}
\includegraphics[width=0.2\textwidth]{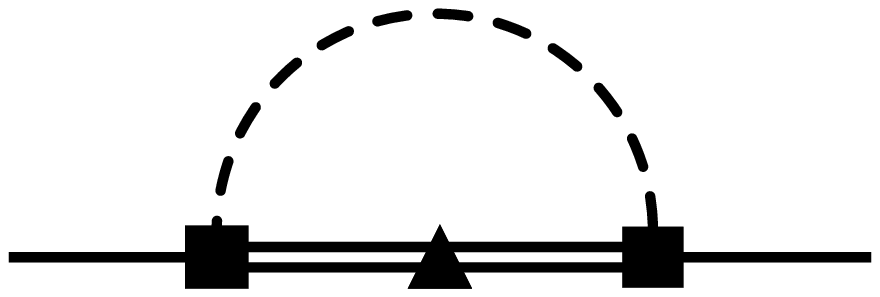}
\hspace{0.05\textwidth}
\includegraphics[width=0.2\textwidth]{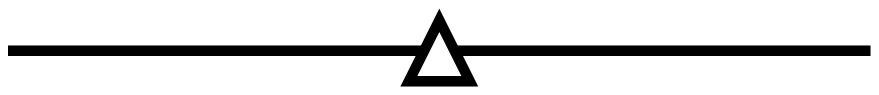}
\caption{Diagrams depicting mass contributions to the nucleons at next-to-next-to-leading order (NNLO)
  in tmHB$\chi$PT in the physical pion basis. The solid, double solid
  and dashed lines denote nucleons, deltas and pions respectively. The
  solid triangle denotes an insertion of the twisted Wilson operator
  as given in~\eqref{E:BLLOphy}. The solid squares denote the coupling
  of the baryons to the axial current whose form is given
  in~\eqref{E:BLTN}. The clear triangle denotes a tree level insertion
  of the operators given in Eqs.~\eqref{E:Namq} and
  \eqref{E:Na2chi}~--~\eqref{E:Nasq}.} 
\label{fig:NNLO}
\end{figure}

The corrections to $M_{N_i}^{(2)}$ come from both the tree level and
the one-loop diagrams, as shown in Fig.~\ref{fig:NNLO}. The twisted
Wilson operator in the free Lagrangian~(\ref{E:BLLOphy}) gives rise to
a tadpole diagram, which produces a contribution of $\c{O}(a
\mathsf{m})$.  The leading Wilson spurions also contribute to $\c{O}(a
\mathsf{m})$ when inserted inside the pion-nucleon loops, and are
partly cancelled by wavefunction corrections. The tree level
contributions come from the operators given in Eqs.~\eqref{E:Namq} and
\eqref{E:Na2chi}~--~\eqref{E:Nasq}. Just as in the untwisted continuum
theory, these act both as the higher dimensional operators and as
counter terms that renormalize divergences from the lower order loop
contributions. For instance, coefficients $n_1^{WM_+}$ and
$n_2^{WM_+}$ are renormalized to absorb divergences from the tadpole
and one-loop contributions mentioned above. As mentioned in
Sec.~\ref{sec:baryons}, the operators with these coefficients are
suppressed by $\L_\chi$ instead of $\L_{\rm QCD}$ because they are the
counterterms for the loop divergences. These coefficients are
taken to be the renormalized coefficients (finite) in the mass
calculations, and to contain the counter terms needed in our
renormalization scheme. The corrections to $M_{N_i}^{(2)}$ read
\begin{align} \label{eq:N2Mass}
\delta M_{N_i}^{(2)}(\omega) &=
 12\,\s_W\,\frac{a\L_{\rm QCD}^2}{\L_\chi^2} \,m_\pi^2
 \log\left(\frac{m_\pi^2}{\L_R^2}\right)\cos(\w)
+16\,g_{\D N}^2\,(\ol{\s}_W - \s_W)\, \frac{a\L_{\rm QCD}^2}{\L_\chi^2}
 \left[\c{J}(m_\pi,\D,\L_R) + m_\pi^2\right]\cos(\w) 
\notag \\ 
&\quad      
-2\left(n_1^{WM_+} + 2\,n_2^{WM_+}\right)
 \frac{a\L_{\rm QCD}^2}{\L_\chi}\,m_q\cos(\w) 
-a^2\Lambda_{\rm QCD}^3\left(b + b_v\right) 
\notag \\ 
&\quad 
+a^2\Lambda_{\rm QCD}^3\left( 2\,n_1^{W_-}\sin^2(\w)
-4\,n_1^{W_+}\cos^2(\omega)\right) \,,
\end{align}
where the function $\c{J}$ is given by
\begin{equation} \label{E:J}
\c{J}(m,\d,\L_R) = 
(m^2 - 2\d^2)\log\left(\frac{m^2}{\L_R^2}\right) +
2\d\sqrt{\d^2 - m^2}\log\left( 
\frac{\d - \sqrt{\d^2 - m^2 + i\varepsilon}}
     {\d + \sqrt{\d^2 - m^2 + i\varepsilon}} \right). 
\end{equation}
Note that the $\c{O}(\varepsilon^4)$ corrections are again the same
for both the proton and the neutron. At maximal twist, the
$\c{O}(\varepsilon^4)$ corrections are given by  
\begin{equation}
\delta M_{N_i}^{(2)}(\w=\pi/2) = 
a^2\L_{\rm QCD}^3\left(2n_1^{W_-} - b - b_v \right) \,,
\end{equation}
while at zero twist, these reduces to the corrections given in
Ref.~\cite{Tib05}. We see that the nucleon masses are also
automatically $\c{O}(a)$ improved at $\c{O}(\varepsilon^4)$. 

To the order we work, the expressions for the nucleon mass corrections
in tmHB$\chi$PT given in Eqs.~\eqref{E:NBLO} and~\eqref{eq:N2Mass},
together with the untwisted continuum HB$\chi$PT expressions for the
nucleon masses, provide the functional form for the dependence of the
nucleon masses on the twist and angle, $\w$, and the quark mass,
$m_q$, which can be used to fit the lattice data.

%
%
%
%
%
%
%
%
%
%
%
%
\subsection{\label{sec:DMassIso} Delta Masses in the Isospin Limit}

Before we present our delta mass expressions, we stress that they can
only be fit to tmLQCD data for sufficiently large quark masses such
that the delta is a stable particle. This corresponds to $m_\pi
\gtrsim$~300~MeV, which is pushing the bounds of validity of 
chiral perturbation theory~\cite{Beane04}. However, these expressions
can be used to study the convergence of $\chi$PT for these pion
masses, where the LECs can be determined. With the value of the LECs
known, the mass calculations can be analytically continued to pion
masses where the delta becomes unstable, and be used to predict
e.g. their lifetimes.  
 
The delta masses in the continuum, infinite volume HB$\chi$PT with a
flavor doublet of degenerate quarks have a similar expansion as that
for the nucleons given in Eq.~\eqref{eq:Nmassexp}. The mass expansion
of the $i^{th}$ delta is conventionally written  
\begin{equation} \label{eq:Tmassexp}
M_{T_i} = M_0\left(\D,\L_R\right) + \D
        + M_{T_i}^{(1)}\left(\D,\L_R\right) 
        + M_{T_i}^{(3/2)}\left(\D,\L_R\right)
        + M_{T_i}^{(2)}\left(\D,\L_R\right) + \ldots
\end{equation}
where
\begin{equation}
T_1 = \D^{++} \,,\qquad T_2 = \D^+ \,,\qquad 
T_3 = \D^0 \,,\qquad  T_4 = \D^- \,,
\end{equation}
and $M_{T_i}^{(n)}$ is the contribution to the $i^{th}$ delta of
$\c{O}(m_q^n)$ calculated in the continuum and infinite volume
two-flavor $\chi$PT in the isospin limit~\cite{TWL05b}. As in the case
of the nucleons, $M_0(\D,\L_R)$ is the renormalized nucleon mass in
the chiral limit, and the parameter, $\D$, is the renormalized mass
splitting between the nucleons and deltas which is independent of the
quark masses.\footnote{There are some subtle issues involved in calculating
  $\D$ in HB$\chi$PT due to the fact that it is a flavor singlet, and
  so can modify all operators/LECs in the chiral Lagrangian. However,
  since one can not vary $\D$, all the LECs associated with modifying
  $\D$ are not determinable. Nevertheless, one can simply fit for the
  physical value of $\D$ determined by the lattice, and use that value
  in all expressions in which it arises. For further discussion on
  this point, see  Refs.~\cite{TWL05b,Tib05PQ}.}   

Both parameters, $M_0$ and $\D$, are flavor singlets, and are
therefore renormalized in the same way. In the nucleon sector, all
flavor singlet mass contributions that are independent of the quark
mass go into renormalizing the parameter $M_0$. In the delta sector,
the choice of which parameter to renormalize, $M_0$ or $\D$, is
arbitrary. For convenience we choose $M_0$ to be the renormalized
nucleon mass in the chiral limit. Thus, $\D$ is simply the difference
between the nucleon and delta masses in the chiral limit, to a given
order in the chiral expansion. Both $M_0$ and $\D$ are parameters
which must be fit to the lattice data. For the delta masses, we use
the same renormalization scheme as for the nucleons.  

We again denote the correction of $\c{O}(\varepsilon^{2n})$ to the
delta masses due to the effects of lattice discretization and twisting
arising from tmLQCD as $\d M_{T_i}^{(n)}$, and the delta masses in
tmHB$\chi$PT are written as
\begin{equation}
M_{T_i}^{tm} = M_0  + \D + (M_{T_i}^{(1)} + \d M_{T_i}^{(1)}) + \ldots
\end{equation}
The leading mass correction arises at tree level from the twisted
Wilson delta operator given in Eq.~\eqref{E:BLLOphy}, 
\begin{equation}\label{eq:DLO}
\d M_{T_i}^{(1)}(\w) = -4\,\ol{\s}_W\,a\,\L_{\rm QCD}^2\cos(\w) \,.
\end{equation}
Just as for the nucleons, this does not split the delta masses and
vanishes at maximal twist. 

\begin{figure}[tbp]
\centering
\includegraphics[width=0.2\textwidth]{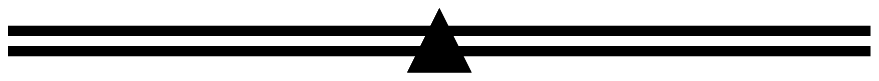}
\hspace{0.05\textwidth}
\includegraphics[width=0.2\textwidth]{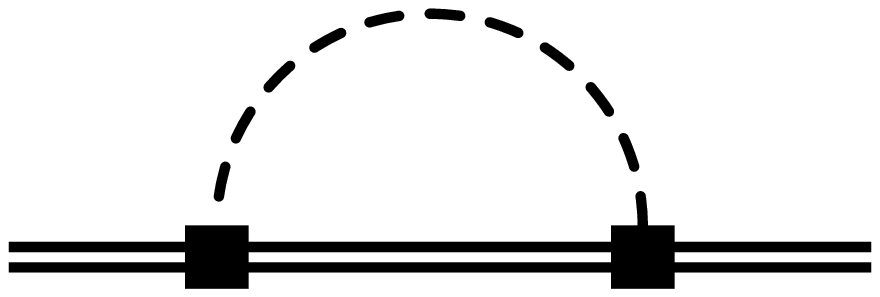}
\hspace{0.05\textwidth}
\includegraphics[width=0.2\textwidth]{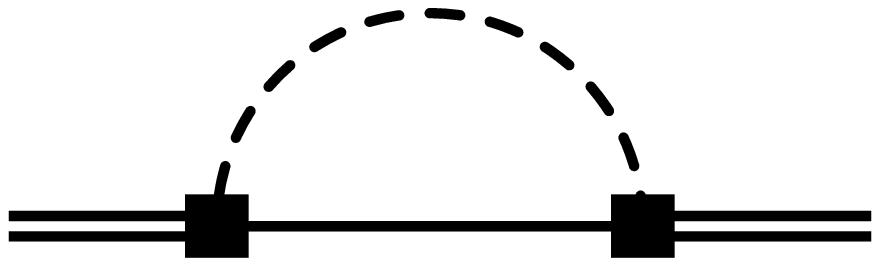}
\caption{Diagrams depicting LO and NLO mass contributions to the
  deltas in tmHB$\chi$PT in the physical pion basis. The solid, double
  solid and dashed lines denote nucleons, deltas and pions
  respectively. The solid triangle is an insertion of the twisted
  Wilson operator given in Eq.~\eqref{E:BLLOphy}. The solid squares
  denote the couplings of the baryons to the axial current whose form
  is given in Eq.~\eqref{E:BLTN}.}  
\label{fig:DLOandNLO}
\end{figure}

\bigskip

The $\c{O}(\varepsilon^3)$ delta mass contributions are similarly
given as for the nucleons.  The contributing diagrams are shown in Figure~\ref{fig:DLOandNLO}.
They do not cause any splitting between the
deltas, and receive no discretization corrections. For completeness,
we list the full mass expression at this order here
\begin{equation}\label{eq:DNLO}
M^{(3/2)}_{T_i} = 
-\frac{25g_{\D\D}^2}{432 \pi f^2}\,m_\pi^3 
-\frac{2g_{\D N}^2}{3(4 \pi f)^2}\,\c{F}(m_\pi,-\D,\L_R).
\end{equation}
For $m_\pi > \D$, the deltas are stable particles, but for 
$m_\pi < \D$, the deltas become unstable (as can be seen from the fact
that the function, $\c{F}(m_\pi,-\D,\L_R)$, picks up an imaginary
component). When the deltas are unstable, one will not be able to use
these expressions to fit the lattice data. However, these expressions
can be used fit the lattice data for $m_\pi \gtrsim$~300~MeV, and then
analytically continued to light enough pion masses for which the
deltas are unstable.   

\begin{figure}[tbp]
\centering
\includegraphics[width=0.2\textwidth]{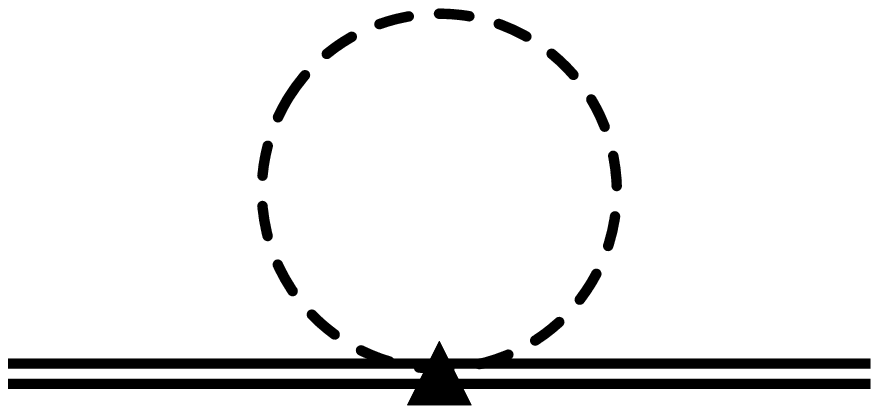}
\hspace{0.05\textwidth}
\includegraphics[width=0.2\textwidth]{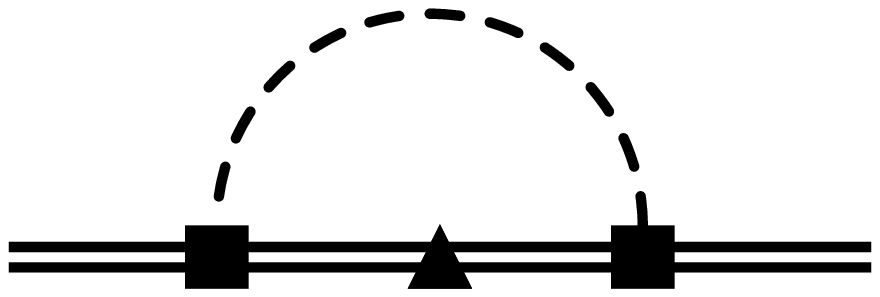}
\hspace{0.05\textwidth}
\includegraphics[width=0.2\textwidth]{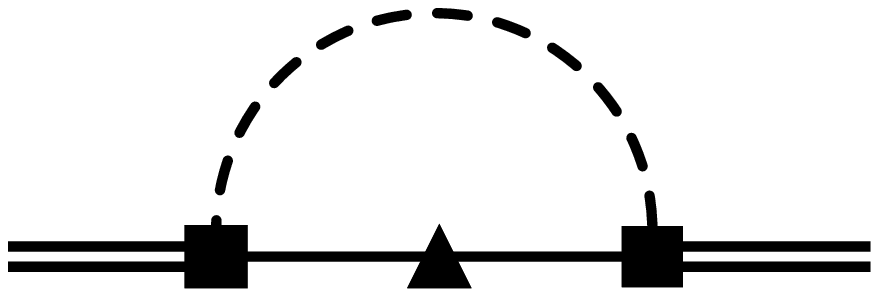}
\hspace{0.05\textwidth}
\includegraphics[width=0.2\textwidth]{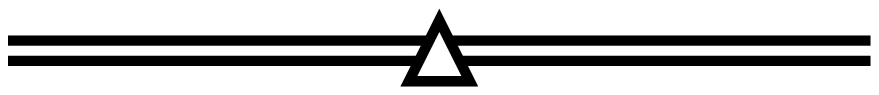}
\caption{Diagrams depicting mass contributions to the deltas at NNLO
  in tmHB$\chi$PT in the physical pion basis. The solid, double solid
  and dashed lines denote nucleons, deltas and pions respectively. The
  solid triangle denotes an insertion of the discretization operator
  in Eq.~\eqref{E:BLLOphy}. The solid squares denote the coupling of
  the baryons to the axial current whose form is given in
  Eq.~\eqref{E:BLTN}. The clear triangle denotes a tree level
  insertion of the operators given in Eqs.~\eqref{E:Tamq}~-~\eqref{E:Na2O4} and
  \eqref{E:Tasq}.} 
\label{fig:DNNLO}
\end{figure}

\bigskip

At $\c{O}(\varepsilon^4)$, contributions due to the effects of
twisting arise from similar diagrams as in the nucleon case, and are
shown in Fig.~\ref{fig:DNNLO}. A splitting in the delta masses first
arises at this order, which comes from the operator with coefficient
$t_2^{W_-}$ given in~\eqref{E:Tasq}. The mass corrections read
\begin{align} \label{eq:DNNLOtau1}
\d M_{T_i}^{(2)}(\w) &= 
 12\,\ol{\s}_W\, \frac{a\L_{\rm QCD}^2}{\L_\chi^2} \,m_\pi^2
 \log\left(\frac{m_\pi^2}{\L_R^2}\right)\cos(\w)
+4\,g_{\D N}^2\,(\ol{\s}_W -\s_W)\,\frac{a\L_{\rm QCD}^2}{\L_\chi^2}\,
 \c{J}(m_\pi,-\D,\L_R)\cos(\w) 
\notag \\
&\quad 
+2\left(t_1^{WM_+} + 2\,t_2^{WM_+}\right)
 \frac{a\L_{\rm QCD}^2}{\L_\chi}m_q\cos(\w)
+a^2\L_{\rm QCD}^3\left(t+t_v\right) 
\notag \\
&\quad 
+a^2\L_{\rm QCD}^3\left(4\,t_1^{W_+}\cos^2(\w)-2\,t_1^{W_-}\sin^2(\w)
 +t_2^{W_-}\d_{T_i}\sin^2(\w)\right) \,,
\end{align}
where 
\begin{equation} \label{E:D2split}
\d_{T_i} =  
\begin{cases}
    -1          &\text{for $T_i = \D^{++}\,,\D^-$} \\
\ \ \frac{1}{3} &\text{for $T_i = \D^+\,,\D^0$}
\end{cases} \,.
\end{equation}
Note the appearance of the mass splitting, $\d_{T_i}$, in $\d
M_{T_i}^{(2)}$. We see from above that starting at
$\c{O}(\varepsilon^4)$, the delta multiplet is split into two 
mass-degenerate pairs, with one pair containing $\D^{++}$ and $\D^-$,
and the other, $\D^+$ and $\D^0$. At maximal twist, $\d M_{T_i}^{(2)}$
becomes 
\begin{equation}
\d M_{T_i}^{(2)}(\w=\pi /2) = 
a^2\L_{\rm QCD}^3\left(t+t_v-2\,t_1^{W_-}+t_2^{W_-}\d_{T_i}\right) \,,
\end{equation}
while at zero twist, it reduces to that given in
Ref.~\cite{Tib05}. Just as in the nucleon case, the delta masses to
$\c{O}(\varepsilon^4)$ are also automatically $\c{O}(a)$ improved. 

As is the case with the nucleons, to the order we work, the
expressions for the delta mass corrections in tmHB$\chi$PT given in
Eqs.~\eqref{eq:DLO} and~\eqref{eq:DNNLOtau1}, together with the
untwisted continuum HB$\chi$PT expressions for the delta masses,
provide the functional form for the dependence of the delta masses on
the twist angle, $\w$, and the quark mass, $m_q$, which can be
used to fit the lattice data.

%
%
%
%
%
%
%
%
%
%
%
%
\subsection{\label{sec:split} Mass splittings}

Having derived the expressions for the nucleon and delta masses in
tmHB$\chi$PT to order $\c{O}(\varepsilon^4)$ in the isospin limit, we
now focus on the mass splittings between the nucleons and between the
deltas. The mass contributions in the continuum, $M_{N_i}^{(n)}$ and
$M_{T_i}^{(n)}$, clearly do not give rise to mass splittings for
the nucleons and deltas, since they are calculated with degenerate
quarks. Therefore, any mass splitting can only come from the mass
corrections arising from tmLQCD. 

From the results of Sec.~\ref{sec:NMassIso} and
Sec.~\ref{sec:DMassIso}, we find that to $\c{O}(\varepsilon^4)$, the
protons and neutrons remain degenerate, while the delta multiplet
splits into two degenerate pairs, with $\D^{++}$ and $\D^-$ in one
pair, and $\D^+$ and $\D^0$ in the other. The splitting between the
degenerate pairs in the delta multiplet is given by  
\begin{equation}
M_{\D^{+,0}} - M_{\D^{++,-}} = 
\frac{4}{3}\,t_2^{W_-}a^2\L_{\rm QCD}^3\sin^2(\w) =
\frac{4}{3}\,t_2^{W_-}a^2\L_{\rm QCD}^3\,\frac{\mu^2}{m_q^2} \,.
\end{equation}
We reiterate here that the $\c{O}(a)$ uncertainty inherent in the
definition of the twist angle results in a correction to 
$M_{\D^{+,0}} - M_{\D^{++,-}}$ of
$\c{O}(a^3)\sim\c{O}(\varepsilon^6)$, which is of higher order than  
we work. Hence to the accuracy we work, we may use $\w_0$ or any other
non-perturbatively determined twist angle for $\w$ above.

Just as the case of the pion mass splitting worked out in
Ref.~\cite{SW04nlo}, this delta splitting must vanish quadratically in
$a\mu = a m_q\sin(\w)$ on general grounds, since the masses do not
violate parity. One would therefore expect, naively, the splitting to
be $\c{O}(a^2m_q^2)\sim\c{O}(\varepsilon^8)$. But as our results show,
there is, in fact, a mass dependence in the denominator such that the
effect is $\c{O}(\varepsilon^4)$. Suppose we take $a^{-1} = 2$~GeV and
$\L_{\rm QCD} = 0.5$~GeV, then we would find a mass splitting 
\begin{equation}
M_{\D^{+,0}} - M_{\D^{++,-}} = 0.04\, t_2^{W_-}\ \text{GeV}.
\end{equation}
Using naive dimensional analysis, we expect $t_2^{W_-} \sim \c{O}(1)$,
and we would have a mass splitting of the delta pairs on the order
of 50~MeV. We point out that in the recent quenched
study~\cite{ARLW05}, a mass splitting of on the order of 50 to 100~MeV
is found. However, our formula can only be applied to lattice data
from unquenched simulations, which have yet to be done, since it is
derived in a fully unquenched theory. Nevertheless, it is encouraging
that the quenched results are not dramatically different from the
estimate we give above with a reasonable value of $\L_{\rm QCD}$.  

Now, the degeneracies we found for the nucleons and the delta
multiplet above hold not only at $\c{O}(\varepsilon^4)$, but in fact
they hold to all orders in tm$\chi$PT. This can be understood by
considering the lattice Wilson-Dirac operator associated with the
action of tmLQCD given in Eq.~(\ref{E:ActionTw}) in the isospin limit with
$\tau_3$-twisting 
\begin{equation}
D_{WD} = \frac{1}{2} \sum_{\mu} \g_\mu (\nabla^\star_\mu + \nabla_\mu) 
        -\frac{r}{2} \sum_{\mu} \nabla^\star_\mu \nabla_\mu 
        + m_0 + i \gamma_5 \tau_3 \mu_0 \,,
\end{equation}
which has the self-adjointness property~\cite{FRLatt04}
\begin{equation}
\t_1 \g_5\,D_{WD}\,\g_5 \t_1 = D_{WD}^\dagger \,.
\end{equation}
It follows then that the propagator for the upper and lower component
of the quark doublet, $\psi_l(x)$, call them $S_u(x,y)$ and
$S_d(x,y)$ respectively, satisfy the relations
\begin{equation}
\g_5\,S_u(x,y)\,\g_5 = S_d^\dagger(y,x) \,,\qquad\qquad
\g_5\,S_d(x,y)\,\g_5 = S_u^\dagger(y,x) \,.
\end{equation}
This means that any baryon two-point correlator which is invariant under the
interchange of the quark states in the quark doublet combined with
hermitian conjugation, leading to the degeneracies mentioned above.
An argument of this type has been given in Ref.~\cite{ARLW05}.  

The same can also be shown in a chiral Lagrangian treatment, as must
be the case. Now one of the symmetries of tmLQCD with two
flavor-degenerate quarks and $\tau_3$-twisting is the pseudo-parity
transformation, $\c{P}^1_F$, where ordinary parity is combined with
a flavor exchange~\cite{FR03},
\begin{equation} \label{E:PF1}
\mathcal{P}^1_F \colon
\begin{cases}
U_0(x) \rightarrow U_0(x_P) \,, \quad x_P = (-\mathbf{x},t) \\
U_k(x) \rightarrow U^\dagger_k(x_P) \,, \quad k = 1,\,2,\,3 \\  
\psi_l(x) \rightarrow i\tau_1 \gamma_0 \psi_l(x_P) \\
\bar{\psi}_l(x) \rightarrow -i\bar{\psi}_l(x_P) \gamma_0 \tau_1
\end{cases} \,,
\end{equation}
where $U_\mu$ are the lattice gauge fields. At the level of
HB$\chi$PT, this is manifested as the invariance of the chiral
Lagrangian under the transformations 
\begin{gather} \label{E:chiP1f}
p(x) \leftrightarrow n(x_P) \,,\qquad 
\D^{++,-}(x) \leftrightarrow \D^{-,++}(x_P) \,,\qquad 
\D^{+,0}(x) \leftrightarrow \D^{0,+}(x_P) \,,\qquad
\otimes^{F}_{k=1}\c{O}_k(x) \rightarrow 
\otimes^{F}_{k=1}\tau_1 \c{O}_k(x_p)\tau_1 \,,
\end{gather}    
where for an operator in the chiral Lagrangian, $\c{O}_k$ is any
operator matrix that contracts with the flavor indices of the the
nucleon ($N$) or delta ($T_\mu$) fields in the operator.  If the $N$ or the $T_\mu$ fields contained in an operator have a total of 2F flavor indices,  $\otimes^{F}_{k=1}\c{O}_k$ is the
tensor product of $F$ operator matrices which contract with the
$F$ distinct pairs of these flavor indices. The degeneracies in
the nucleons and the delta multiplets discussed above would then follow
if all the operators in the chiral Lagrangian that contribute to the
baryon masses have a structure that satisfies the condition 
\begin{equation} \label{E:symC}
\otimes^{F}_{k=1}\c{O}_k(x)
= \otimes^{F}_{k=1}\tau_1\c{O}_k(x_P)\tau_1 \,,
\end{equation}

Consider first the case for the nucleons. Since the nucleon fields are
vectors in flavor space, we can take $F=1$ without loss of
generality (the nucleon fields can only couple to one operator
matrix). Since the chiral Lagrangian is built with just $\c{M}_\pm$, 
$\c{W}^{tw}_\pm$, $\c{A}_\mu$, and $\c{V}_\mu$, the operator matrix
$\c{O}_k$ can only be constructed from combinations of these
fields. We need not consider combinations involving just $\c{M}_+$ and
$\c{W}^{tw}_+$, since the flavor structure of both is trivial,
i.e. proportional to the identity. We also need not consider mass
contributions arising from pion loops, because they must have the same
flavor structure as the tree level local counterterms used to cancel
the divergences in these loops. Therefore, we do not have to consider
operators involving $\c{A}_\mu$ and $\c{V}_\mu$, which give rise to
mass contributions only through pion-nucleon interactions. This leaves
us with only combinations involving $\c{M}_-$ and $\c{W}^{tw}_-$ as
possible candidates to break the degeneracy in the nucleons. As was
discussed in Sec.~\ref{sec:baryons}, because of the parity-flavor
symmetry of tmLQCD, $\c{O}_k$ can not contain just a single
$\c{M}_-$ or $\c{W}^{tw}_-$, but must always have an even number from
the set $\{\c{M}_-,\c{W}^{tw}_-\}$. Now any such combination would
indeed have a pure tree level part, however, it is also proportional
to the identity in flavor space. Thus there is no operator matrix,
$\c{O}_k$, that one can construct which violates the condition  
$\c{O}_k(x) = \tau_1\c{O}_k(x_P)\tau_1$.

The arguments for the case of the deltas runs similar to that for the
nucleons. For the same reason given in the nucleon case, we need not
consider operator structures that involve $\c{M}_+$, $\c{W}^{tw}_+$,
$\c{A}_\mu$, and $\c{V}_\mu$. We need only consider operator
structures involving an even number from the set
$\{\c{M}_-,\c{W}^{tw}_-\}$. For the deltas, $F$ can be three since
each delta field has three flavor indices. But since two of the
$\c{O}_k$ in $\otimes^{3}_{k=1}\c{O}_k$ must come from the set
$\{\c{M}_-,\c{W}^{tw}_-\}$ to satisfy the parity-flavor symmetry of
tmLQCD, we can take $F$ to be at most two without loss of
generality. Now each of $\c{M}_-$ and $\c{W}^{tw}_-$ has a tree level
part that is proportional to $\tau_3$, thus, under $\c{P}^1_F$,
$\c{O}_1\otimes\c{O}_2$ where $\c{O}_k$ can be either $\c{M}_-$ or
$\c{W}^{tw}_-$, satisfies the symmetry condition
(\ref{E:symC}). Therefore, one can not construct operators for the
deltas that break the degeneracy between the pairs in the delta
multiplet.

%
%
%
%
%
%
%
%
%
%
%
%
\section{\label{sec:mass} Nucleon and delta masses away from the
  isospin limit}   

In this section, we present results for mass corrections due to
twisting away from the isospin limit, where the quarks are now mass
non-degenerate. To the order we work, the corrections due to the mass
splitting come in only at tree level. For clarity, we will only point
out the change arising from the quark mass splitting; we will not
repeat the discussion on the nucleon and delta masses that are the
same both in and away from the isospin limit.

\subsection{\label{sec:Tdiag} The flavor-diagonal basis for the mass
  matrix at $\mathcal{O}(\varepsilon^4)$} 

The natural choice for splitting the quark doublet is
to use the real and flavor-diagonal Pauli matrix, $\tau_3$, since the
quark states one uses on the lattice correspond to the quarks in QCD
in the continuum limit. But as was discussed in Sec.~\ref{sec:EContL}
above, twisting can not be implemented with $\tau_3$ in this case (the
fermionic determinant would be complex otherwise), and so $\tau_1$
is used instead. This means that the quark mass matrix in
$\mathcal{L}_{\rm eff}$ given in Eq.~(\ref{E:CLeff}), 
$m + i\mu\gamma_5\tau_1 - \epsilon_q\tau_3$, can never be made
flavor-diagonal through an appropriate change of basis if both the 
twist and the mass splitting are non-vanishing, because $\tau_1$
and $\tau_3$ can not be simultaneously diagonalized.

Since the twist is implemented by a flavor nondiagonal Pauli matrix
away from the isospin limit, flavor mixings are induced for
non-zero twist: the quark states in tmLQCD are now linear
combinations of the physical quarks of continuum QCD. At the level of
the chiral effective theory, this manifests itself in that the
hadronic states described by the tm$\chi$PT Lagrangian are linear
combinations of the continuum QCD hadronic states we observe, viz. the
pions, nucleons, deltas, etc. 

If the effects from twisting are perturbative as compared to the
isospin breaking effects, the hadronic states described by tm$\chi$PT
will be ``perturbatively close'' to their corresponding continuum QCD
states, i.e. the difference between them is small compared to the
scales in the theory (see Appendix~\ref{sec:appDDM} for an explicit
demonstration). In this case, we can still extract QCD observables
directly from tm$\chi$PT, as the corrections will be perturbative in
the small expansion parameter. However, if the twisting effects are on
the same order as the isospin breaking effects so that the flavor
mixings are large, these corrections will not be perturbative.%
\footnote{A qualitative guide to the size of the flavor
  mixings can be found in the ratios of two-point correlation
  functions. Define the ratio of QCD delta states by 
  \begin{equation*}
  R_{ij} \equiv 
  \frac{\langle \D^i \ \D^j \rangle + \langle \D^j \ \D^i \rangle}
  {\langle \D^i \ \D^i \rangle + \langle \D^j \ \D^j \rangle} \,.
  \end{equation*}
  Flavor mixing should be small if the off diagonal elements of
  $R_{ij}$ are small. To determine the size of the flavor mixings
  quantitatively, one has to look at the splitting in the delta
  multiplet. We will discuss further in the text below.} 
Nevertheless, one can still extract information for the QCD
observables: One can still measure the masses of these tmLQCD hadronic
states in lattice simulations, and one can fit these to the analytic
expressions for these masses calculated in tm$\chi$PT to extract the
values of the LECs. The LECs associated with the continuum $\chi$PT
contributions have the same numerical values as in
tm$\chi$PT. Therefore, if one determines these from tmLQCD
simulations, one knows the masses of the QCD hadronic states. 

At the order we work, flavor mixings are manifested in the appearance
of flavor non-conserving pion-baryon vertices in the Feynman rules of
tmHB$\chi$PT, and in that the baryon mass matrix is not
flavor-diagonal. Since we work in the physical pion basis where the
twist is carried by the Wilson spurion (now flavor non-diagonal)
instead of the mass spurion (now flavor-diagonal), flavor mixings can
only arise from operators with one or more insertions of the Wilson
spurion field. Because of this, the flavor non-conserving pion-baryon
vertices and the non-diagonal terms in the mass matrix must be
proportional to $a$, the lattice spacing, and so must vanish in the
continuum limit where the effects of the twist are fake and can be
removed by a suitable chiral change of
variables~\cite{FetalLatt99,Fetal01}.\footnote{This shows again the 
  convenience of the pion physical basis, where all the effects of
  symmetry breaking in the lattice theory are parametrized and
  contained in the Wilson spurion fields, which then vanish as the
  symmetries are restored in the continuum limit.} 

For the nucleons, flavor mixings induce only unphysical flavor
non-conserving pion-nucleon vertices which vanish in the continuum
limit; the nucleon matrix is still flavor diagonal at the order we
work. In fact, this is true to all orders in tmHB$\chi$PT. The reason
is the same as that given in Sec.~\ref{sec:split}. We need only
consider the tree level part of the possible operator structures that
one can construct from the spurion fields in tmHB$\chi$PT. Now the
only spurion field that has a tree level part with non-diagonal
flavor structure is $\c{W}^{tw}_-$, and as we discussed above, it must
be paired either with another $\c{W}^{tw}_-$ or with $\c{M}_-$, which
renders the flavor structure of the tree level part of the combination
trivial. Thus we may take the basis of nucleons used in the
tmHB$\chi$PT Lagrangian as the physical nucleon basis.  

For the deltas, not only are there flavor non-conserving pion-delta
vertices, at the order we work, the delta mass matrix is already 
flavor nondiagonal at tree level. This happens for the deltas
because the tensor nature of the $T_\mu$ field allows more freedom in
the way the flavor structure of the delta operator can be
constructed. Thus, in order to have only physical tree level mass
terms for the deltas, we must change to a basis where the delta mass
matrix is diagonal, which can now only be done order by order. 

When diagonalizing the delta mass matrix, we need, in principle, to
diagonalize the mass matrix that contains all the mass contributions
from both tree and loop level to the order that one works. But we find
the difference between diagonalizing the delta mass matrix including
both tree and loop level contributions at the order we work, and
diagonalizing that with only the tree level mass contributions, give
rise to corrections only to the loop level mass contributions, which
are higher order than we work. Thus, we will diagonalize the delta
mass matrix containing just the tree level mass terms in our
calculation for the delta masses. 

To the order we work, if the tree level mass is given by 
\begin{equation} \label{E:Tbasis}
v_{\bar{\Delta}} M_\Delta v_\Delta \,, \qquad
v_{\bar{\Delta}} = 
\begin{pmatrix}
\bar{\Delta}^{++} & \bar{\Delta}^0 & \bar{\Delta}^+ & \bar{\Delta}^-
\end{pmatrix} \,, \qquad
v_\Delta = 
\begin{pmatrix}
\Delta^{++} & \Delta^0 & \Delta^+ & \Delta^-
\end{pmatrix}\transpose \,, 
\end{equation}
where $v_{\bar{\Delta}}$ and $v_\Delta$ are vectors of the delta
basis states used in the tmHB$\chi$PT Lagrangian, and $M_\Delta$ is
the tree level mass matrix, the physical delta basis is given by  
\begin{equation} \label{E:newT}
v'_\Delta = S^{-1} \cdot v_\Delta \,,\qquad 
v'_{\bar{\Delta}} = v_{\bar{\Delta}} \cdot S \,, 
\end{equation}
where $S$ is the matrix of eigenvectors of $M_\Delta$ such that
\begin{equation}
S \cdot M_\Delta \cdot S^{-1} = \mathcal{D} \,,
\end{equation}
with $\c{D}$ the corresponding diagonal eigenvalue matrix. This
implies that
\begin{equation}
v_{\bar{\Delta}} M_\Delta v_\Delta =
(v'_{\bar{\Delta}} \cdot S^{-1}) \cdot
(S \cdot \c{D} \cdot S^{-1}) \cdot (S \cdot v'_\Delta) = 
v'_{\bar{\Delta}} \c{D} \, v'_\Delta \,.
\end{equation} 
The full details of the diagonalization are provided in
Appendix~\ref{sec:appDDM}. In the following sections, we will work in
this basis for calculating the delta masses.

The Feynman rules in the new basis are obtained from the same
tmHB$\chi$PT Lagrangian given above in Sec.~\ref{sec:baryons} but with
each of the delta flavor states now rewritten in terms of the new 
delta flavor states given by the defining relations
(\ref{E:newT}). Note that changing to the new delta basis induces new 
unphysical flavor non-conserving vertices in the delta interaction
terms given in (\ref{E:BLTN}), because in terms of the new basis
states, flavors are mixed. However, these flavor mixing components are
proportional to the off-diagonal elements of $S$, which are
proportional to the lattice spacing as well as the twist angle (see
Appendix~\ref{sec:appDDM}). Thus they vanish in the limit of vanishing
twist or lattice spacing, and so the unphysical vertices arising from
them also vanish in these limits. 

We note and reiterate here that in the isospin limit, this order by
order mass matrix diagonalization is unnecessary as one can always
rotate to a basis where the twist is flavor-diagonal from the outset,
and issues of flavor nonconserving vertices and non-diagonal mass matrices
due to flavor mixings do not arise.\footnote{In fact, as is shown
  in Appendix~\ref{sec:appDDM}, if one insists on remaining in the basis
  where the twist in flavor non-diagonal, one would find that the
  unphysical terms arising from flavor mixings do not vanish in the
  continuum limit.} 

%
%
%
%
%
%
%
%
%
%
%
%
\subsection{\label{sec:tmNMass} The nucleon masses}

Away from the isospin limit, the first change caused by the mass
splitting  occurs in the continuum mass contribution $M_{N_i}^{(1)}$,
since the quark masses  
\begin{equation}
m_u = m_q - \epsilon_q \,,\qquad\qquad m_d = m_q + \epsilon_q 
\,,\qquad\qquad m_q\;,\epsilon_q > 0 \,,
\end{equation}  
are no longer equal.

At the order we work, the only other change due to the mass splitting
appears at $\c{O}(\varepsilon^4)$ in the contribution from the
$\c{O}(a\mathsf{m})$ nucleon operator with coefficient $n^{WM_+}$
given in Eq.~(\ref{E:Namq}). In the isospin limit, its contribution to
$\d M_{N_i}^{(2)}(\w)$ is proportional to $m_q$, but away from the
isospin limit, it becomes 
\begin{equation} \label{E:Nchange}
2\,n_1^{WM_+}a\,\Lambda_\chi\,m_q\cos(\w) \longrightarrow
2\,n_1^{WM_+}a\,\Lambda_\chi\,m_i\cos(\w) \,,
\end{equation} 
where
\begin{equation} \label{E:muddef}
m_i = 
\begin{cases}
m_u &\text{for $i=p$} \\
m_d &\text{for $i=n$} 
\end{cases} \;.
\end{equation}
The corrections to the nucleon masses from the effects of lattice
discretization and twisting are otherwise the same as those given in 
Eqs.~\eqref{E:NBLO} and~\eqref{eq:N2Mass}. 

Note that the nucleon masses are automatically $\c{O}(a)$ improved,
just as in the isospin limit.

%
%
%
%
%
%
%
%
%
%
%
%
\subsection{\label{sec:tmDmass} The delta masses} 

Away from the isospin limit ($\epsilon_q \neq 0$), we calculate the
delta mass and mass corrections in the basis where the delta mass
matrix is diagonal to the order we work. This diagonalization is
worked out in Appendix~\ref{sec:appDDM}, where we have obtained general
expressions for the new delta basis that are valid in the range from
$\epsilon_q = 0$ to $\epsilon_q \sim a\L_{\rm QCD}^2$. Here, we present
the case where $\epsilon_q > 0$ and $\epsilon_q \sim m_q \sim a\L_{\rm
  QCD}^2 \gg a^2\L_{\rm QCD}^3$, which is a regime that simulations
in the near future can probe. To the order we work, we may take the
new delta basis states in this regime to be   
\begin{align}
T'_1 \leftrightarrow 
| \D_1 \rangle &= \c{C}_1 \left[ 
| \D^{++} \rangle + \frac{\sqrt{3} B}{4 A} | \D^0 \rangle \right] 
\,,\qquad &
T'_3 \leftrightarrow 
| \D_3 \rangle &= \c{C}_3 \left[ 
\left( 1 + \frac{B}{4A} \right) | \D^{0} \rangle 
- \frac{\sqrt{3} B}{4 A} | \D^{++} \rangle \right] \,,
\notag\\
T'_2 \leftrightarrow 
| \D_2 \rangle &= \c{C}_2 \left[ 
| \D^{+} \rangle + \frac{\sqrt{3} B}{4 A} | \D^- \rangle \right]
\,,\qquad &
T'_4 \leftrightarrow 
| \D_4 \rangle &= \c{C}_4 \left[ 
\left( 1 - \frac{B}{4A} \right) | \D^{-} \rangle 
- \frac{\sqrt{3} B}{4 A} | \D^+ \rangle \right] \,,
\end{align}
where $T'_i = \D_i$ denote the deltas in the new basis, $\c{C}_i$
are normalization factors, and 
\begin{equation}\label{E:AandB}
A = 2\,\e_q\, \left( \g_M + t_1^{WM_+} 
    \frac{a \L_{QCD}^2}{\L_\chi}\cos(\w) \right) \,,\qquad 
B = t_2^{W_-}a^2\L_{QCD}^3\sin^2(\w) \,.
\end{equation}
Note that $A\sim\c{O}(\varepsilon^2)$ and $B\sim\c{O}(\varepsilon^4)$
in our power counting, so $B/A \sim \c{O}(\varepsilon^2)$ and the
effects of the flavor mixings is perturbative.

The masses of these states are comprised of the continuum expressions
given in Ref.~\cite{TWL05b} and corrections due to the effects of
discretization and twisting. Note the continuum expressions for the
delta masses here are necessarily changed from that in the isospin limit
because $m_u \neq m_d$. The mass corrections due to the effects of
lattice discretization and twisting, come in at tree level; the loop
contributions remain unchanged from that in the isospin limit. The
tree level mass contributions to the order we work have been worked
out in  Eq.~\eqref{E:Dmtx}, in the process of diagonalization. We list
here the full delta mass corrections to $\c{O}(\varepsilon^4)$, which
we denote by $\d M_i$, to the mass of the delta state denoted by
$T'_i$: 
\begin{align} \label{E:TMassC}
\d M_i(\w) &= 
-4\,\ol{\s}_W\,a\,\L_{QCD}^2\cos(\w)
\notag \\
&\quad
+12\,\ol{\s}_W\frac{a \L_{\rm QCD}^2}{\L_\chi^2}\,m_\pi^2
 \log\left(\frac{m_\pi^2}{\L_R^2}\right)\cos(\w)
+4\,g_{\D N}^2\,(\ol{\s}_W -\s_W)\frac{a \L_{\rm QCD}^2}{\L_\chi^2}
 \,\c{J}(m_\pi,-\D,\L_R)\cos(\w)
\notag\\
&\quad
+2\,t_1^{WM_+}\frac{a \L_{\rm QCD}^2}{\L_\chi}\frac{m'_i}{3}\cos(\w) 
+4\,t_2^{WM_+}\frac{a \L_{\rm QCD}^2}{\L_\chi}m_q\cos(\w) 
\notag \\
&\quad
+a^2\L_{QCD}^3(t +t_v) 
+a^2\L_{QCD}^3\left(4\,t_1^{W_+}\cos^2(\w)-2\,t_1^{W_-}\sin^2(\w)
+t_2^{W_-}\d'_i\sin^2(\w)\right) \,,\qquad i=1\,,\ldots,4 \,,
\end{align}
where
\begin{equation}
m'_i = 
\begin{cases}
3m_u      &\text{for $i=1$} \\
2m_u+m_d  &\text{for $i=2$} \\
 m_u+2m_d &\text{for $i=3$} \\
3m_d      &\text{for $i=4$}
\end{cases} \;,\qquad\qquad
\d'_i = 
\begin{cases}
\ \ 0        &\text{for $i=1\,,\,4$} \\
-\frac{2}{3} &\text{for $i=2\,,\,3$} \\
\end{cases} \;.
\end{equation}
Note that $\d M_i(\w)$ as given in Eq.~\eqref{E:TMassC}, is the same
as the sum of $\d M_{T_i}^{(1)}$ and $\d M_{T_i}^{(2)}$ as given in 
Eqs.~\eqref{eq:DLO} and~\eqref{eq:DNNLOtau1} respectively, but with
the changes 
\begin{equation} \label{E:Tchange}
2\,t_1^{WM_+}\frac{a\L_{\rm QCD}^2}{\L_\chi}m_q\cos(\w)
\longrightarrow 
2\,t_1^{WM_+}\frac{a\L_{\rm QCD}^2}{\L_\chi}\frac{m'_i}{3}\cos(\w)
\,, \qquad
\d_{T_i} \longrightarrow \d'_i \,.
\end{equation} 

The full expressions for the delta masses can be obtained when the
continuum contributions are included. To the order we work, one  
can obtain the complete mass expression for delta denoted by $T'_i$ to
$\c{O}(\varepsilon^4)$ in tmHB$\chi$PT by adding the mass corrections,
$\d M_i(\w)$, to the continuum mass of the delta denoted by $T_i$,
whose expression can be found in Ref.~\cite{TWL05b}.  

We stress here that one can not take the isospin limit from any of the 
expressions give above in this subsection. They have been derived for
$\epsilon_q \neq 0$ and with the assumption that the twisting effects
are much smaller than the isospin breaking effects. One must use the
general formulae given in Eqs.~\eqref{E:Smtx} and \eqref{E:Dmtx} when
considering cases where these conditions are not true.

Observe that away from the isospin limit, the delta masses are also
automatically $\c{O}(a)$ improved at maximal twist ($\w = \pi/2$), as
all terms proportional to $a$ in $\d M_i$ are proportional to
$\cos(\w)$ as well. Hence, to the order we work, the contributions due
to the isospin breaking are the same as that in the continuum at maximal
twist.

%
%
%
%
%
%
%
%
%
%
%
%
\section{\label{sec:conc} Summary}

In this paper we have studied the mass spectrum of the nucleons and
the deltas in tmLQCD with mass non-degenerate quarks using effective
field theory methods. We have extended heavy baryon chiral
perturbation theory for $SU(2)$ to include the effects of the twisted
mass, and we have done so to $\c{O}(\varepsilon^4)$ in our power
counting, which includes operators of
$\c{O}(a\mathsf{m}^2,ap^2,a^2)$. Using the resulting tmHB$\chi$PT, we
have calculated the nucleon and the delta masses to
$\c{O}(\varepsilon^4)$, and we found them to be automatically
$\c{O}(a)$ improved as expected from the properties of tmLQCD.

Because of the twisting, the vacuum is no longer aligned with the
identity in flavor space, which has non-trivial effects on the
physical excitations (pions) of the theory. Also, depending on whether
the quarks are mass degenerate or not, the way twisting is implemented
determines what the physical baryon states are in the theory. We have
highlighted these subtleties when doing calculations in
tmHB$\chi$PT. 

In order for the pions in the theory to be physical, we have to make a
particular (non-anomalous) chiral change of variables to undo the
twisting effects. This requires the knowledge of the twisting angle,
but once that is determined, the physical pion basis can be determined
{\it \`a priori}. However, whether or not the nucleons and deltas are
physical must still be determined from the theory. In the isospin
limit, both the nucleon and the delta mass matrices are diagonal, and
so the nucleon and delta states contained in the $N$ and $T_\mu$
fields are physical. However, away from the isospin limit, only the
nucleon mass matrix remains diagonal. Thus, the $N$ field can still be
regarded as physical, but the physical deltas are now linear
combinations of the flavor states contained in the $T_\mu$ field. This
can be understood from the fact the at the quark level, the physical
QCD states, the $u$ and $d$ quarks, are eigenstates of $\tau_3$ but
not of $\tau_1$. So only in the isospin limit, where the twist can
always be implemented by the flavor-diagonal Pauli matrix, $\tau_3$,
are the states contained in the quark doublet physical quarks. Away
from the isospin limit, the twist can not be implemented by $\tau_3$
anymore, and the eigenstates of the Hamiltonian of the theory are
composed of linear combinations of the $u$ and $d$ quarks. 

The physical states in tm$\chi$PT are in general a mixture of those in
the untwisted $\chi$PT. The size of the mixture is determined by the
relative sizes of the discretization effects, which are $\c{O}(a^2)$,
and the isospin splitting effects, which are $\c{O}(\epsilon_q)$. In
this work, we have given general expressions for the nucleon and delta
masses with respect to this mixing of states that are valid in the
range from $\epsilon_q = 0$ to $\epsilon_q \sim a\L_{\rm QCD}$.

The quantities which provided the motivation for this work and turned
out to be most interesting are the mass splittings between the
nucleons and between the deltas. We found that in the isospin limit,
the nucleon masses do not split to any order in tm$\chi$PT, while the
delta multiplet splits into two degenerate pairs. This can be
understood from the symmetries of tmLQCD at the quark level, and as we
have shown, also at the level of tm$\chi$PT. The mass splitting between
the multiplets, $M_{\D^{+,0}} - M_{\D^{++,-}}$, first arises from a
tree level contribution at $\c{O}(a^2)$, and it gives an indication of
the size of the flavor breaking in tmLQCD. This splitting in the delta
multiplet will be easier to calculate in lattice simulations than the
corresponding quantity $m_{\pi_3}^2 - m_{\pi_{1,2}}^2$ in the meson
sector~\cite{SW04nlo}, since it involves no quark disconnected
diagrams. 

Twisted mass HB$\chi$PT can also be extended to partially quenched
theories (such extension of tm$\chi$PT for pions has recenly been
done~\cite{MSS04}). This will be useful in the near future as
numerical studies of tmLQCD move from quenched simulations to the more
realistic, if more computationally demanding, partially quenched
simulations.

\section*{Acknowledgments}
We thank Stephen Sharpe for many useful discussions, and Brian Tiburzi
for helpful comments. This research was supported by the
U.S. Department of Energy Grant Nos. DE-FG02-96ER40956 (J.M.S.W.), and
DE-FG03-97ER41014 (A.W.-L.).

%
%
%
%
%
%
%
%
%
%
%
%
\appendix

\section{\label{sec:appD5SB} Absence of additional dimension five
  symmetry breaking operators induced by the mass splitting}

In this appendix, we show that the mass splitting does not induce any
symmetry breaking terms in the effective continuum Lagrangian at the
quark level at quadratic order. The (mass) dimension six operators in
the Symanzik Lagrangian we can drop for the same reason given in
Ref.~\cite{SW04}, since they are either of too high order (cubic or
higher in our expansion) or they do not break the symmetries further
than those of lower dimensions. For dimension five operators, we will
show that the only allowable terms by the symmetries of the lattice
theory are those that either vanish by the equations of motion, or can
be removed by suitable $\mathcal{O}(a)$ redefinitions of the parameters 
in $\mathcal{L}_0$, the effective Lagrangian in the continuum limit
(the lowest order effective Lagrangian). 

In the mass-degenerate case~\cite{SW04}, the only dimension five
operator that appears is the Pauli term. Since in the limit of
vanishing mass splitting (the isospin limit) the mass non-degenerate
theory must be the same as the mass-degenerate theory, any additional
operators induced by the mass splitting must be proportional to the mass
splitting. These can only be of the form
\begin{align} \label{E:dim5op}
&\epsilon_q^2 \bar{\psi} \mathcal{O}_0 \psi: \quad
\mathcal{O}_0 = \Gamma_0\backslash\{\mathbb{1}\}\,, \quad
\Gamma_0 = \{\mathbb{1} \,,\, \tau_k \,,\, \gamma_5 \,,\,\gamma_5\tau_k\}
\,, \quad k = 1,\,2,\,3,  
&{\rm dim}\,[\mathcal{O}_0] = 0 \,, \notag \\ 
&\epsilon_q \bar{\psi} \mathcal{O}_1 \psi: \quad
\mathcal{O}_1 =
\{D\!\!\!\!/\,\Gamma_0\,,\,m\,\Gamma_0\,,\,\mu\,\Gamma_0\,\} 
\backslash\{D\!\!\!\!/\;\tau_3\,,\,m\tau_3\}\,, 
&{\rm dim}\,[\mathcal{O}_1] = 1 \,,
\end{align}
where the notation ``$P \backslash Q$'' means ``the set P excluding
the set Q''. The quantities $\mathcal{O}_0$ and $\mathcal{O}_1$ are
all the possible independent structures with the correct dimension,
which do not lead to dimension five operators vanishing by the
equations of motion, or are not removable by redefinitions of parameters
in $\mathcal{L}_0$. However, none of these operators are allowed under
the symmetries of the lattice theory. Specifically, they are forbidden
by charge conjugation ($\mathcal{C}$) and the pseudo-parity
transformations that combine the ordinary parity transformation
($\mathcal{P}$) with a parameter sign change
\begin{equation}
\widetilde{\mathcal{P}} \equiv 
\mathcal{P} \times (\mu \rightarrow -\mu) \,,
\end{equation}
or a flavor exchange or both
\begin{equation} \label{E:pseudoP}
\mathcal{P}^2_{F,\,\epsilon_q} \equiv
\mathcal{P}^2_F \times (\epsilon_q \rightarrow -\epsilon_q) \,,\quad 
\mathcal{P}^3_F \,,
\end{equation}
where
\begin{equation} \label{E:PF23}
\mathcal{P}^{2,3}_F \colon
\begin{cases}
U_0(x) \rightarrow U_0(x_P) \,, \quad x_P = (-\mathbf{x},t) \\
U_k(x) \rightarrow U^\dagger_k(x_P) \,, \quad k = 1,\,2,\,3 \\  
\psi(x) \rightarrow i\tau_{2,3} \gamma_0 \psi(x_P) \\
\bar{\psi}(x) \rightarrow -i\bar{\psi}(x_P) \gamma_0 \tau_{2,3}
\end{cases} \,,
\end{equation}
and $U_\mu$ are the lattice link fields.  Note that we have displayed the symmetries of the lattice 
theory~\cite{FR03,FRLatt04} in the form which applies to the effective
continuum theory.  

In Table~1, we show explicitly which symmetry forbids each of the
possible structures of $\mathcal{O}_0$ and $\mathcal{O}_1$ listed
in~(\ref{E:dim5op}).\footnote{Most of what we show can be readily
  inferred from~\cite{FRLatt04}. What is new here is the need for
  $\widetilde{\mathcal{P}}$, and the use of
  $\mathcal{P}^{2}_{F,\,\epsilon_q}$.} We group the operators in
columns according to the symmetry under which they are forbidden. 
\begin{table}[tbp] \label{T:OpStr}
\begin{tabular}{|c|c|c|c|c|}
\hline \hline
Structure & 
\hspace{1.5cm}$\mathcal{C}$\hspace{1.5cm} & 
\hspace{1.5cm}$\mathcal{P}^3_F$\hspace{1.5cm} & 
\hspace{0.8cm}$\widetilde{\mathcal{P}}$\hspace{0.8cm} &
\hspace{0.5cm}$\mathcal{P}^2_{F,\,\epsilon_q}$\hspace{0.5cm}
\\ \hline
$\c{O}_0$ & $\tau_2$, $\gamma_5\tau_2$ & 
$\tau_1$, $\gamma_5$, $\gamma_5\tau_3$ & $\gamma_5\tau_1$ & $\tau_3$
\\ \hline
$\c{O}_1$ & $D\!\!\!\!/\;\gamma_5\times\{\mathbb{1},\,\tau_1,\,\tau_3\}$ &
$D\!\!\!\!/\;\tau_1$, $D\!\!\!\!/\;\tau_2$ &
$D\!\!\!\!/\;\gamma_5\tau_2$ & \\ 
& $m\tau_2$, $m\gamma_5\tau_2$ & 
$m\times\{\tau_1,\,\gamma_5,\,\gamma_5\tau_3\}$ & $m\gamma_5\tau_1$ &\\
& $i\mu\tau_2$, $i\mu\gamma_5\tau_2$ &
$i\mu\times\{\tau_1,\,\gamma_5,\,\gamma_5\tau_3\}$ & $i\mu\tau_3$ &
$i\mu\gamma_5\tau_1$ 
\\ \hline \hline 
\end{tabular}
\caption{The structures of the dimension five operators that are
  non-vanishing by the equations of motion and non-removable by 
  parameter redefinitions. They are classified by the symmetries that
  forbid them.}  
\end{table}

The conclusion of the above discussion is that the mass splitting does
not induce any additional operators that do not vanish by the
equations of motion, or can not be removed by redefinitions of the
parameters in the theory. Thus beyond $\mathcal{L}_0$, the effective
continuum Lagrangian contains only the Pauli term to the order we
work, exactly as in the mass-degenerate case.

%
%
%
%
%
%
%
%
%
%
%
%
\section{\label{sec:appDDM} Diagonalization of the delta mass matrix}

Here we diagonalize the tree level mass matrix for the delta states.
We reiterate that the difference between first diagonalizing the tree
level mass contributions, then calculating loop effects, versus
calculating the loop contributions then diagonalizing, is of higher
order than we work. To proceed, first we list all the independent
operators to $\c{O}(\varepsilon^4)$ that have tree level mass
contributions,  
\begin{align} \label{E:Mtree}
\c{O}(\varepsilon^2): \quad &
(\ol{T}_\mu \c{M}_+ T_\mu) \,, \quad 
(\ol{T}_\mu T_\mu)\,{\rm tr}(\c{M}_+) \,, \quad 
(\ol{T}_\mu T_\mu)\,{\rm tr}(\c{W}^{tw}_+) \notag \\
\c{O}(a\mathsf{m}): \quad &
(\ol{T}_\mu \c{M}_+ T_\mu)\,{\rm tr}(\c{W}^{tw}_+) \,, \quad  
(\ol{T}_\mu T_\mu)\, {\rm tr}(\c{W}^{tw}_+)\,{\rm tr}(\c{M}_+)  \notag \\
\c{O}(a^2): \quad &
(\ol{T}_\mu T_\mu)\,
{\rm tr}(\c{W}^{tw}_+)\,{\rm tr}(\c{W}^{tw}_+) \,, \quad
(\ol{T}_\mu T_\mu)\,
{\rm tr}(\c{W}^{tw}_- \c{W}^{tw}_-) \,, \quad
\ol{T}^{kji}_\mu(\c{W}^{tw}_-)^{ii'}(\c{W}^{tw}_-)^{jj'} T^{i'j'k}_\mu
\end{align} 
The tree level delta mass matrix at the order we work, $M_\Delta$, is
then given by 
\begin{align}
v_{\bar{\Delta}} M_\Delta v_\Delta &=
v_{\bar{\Delta}} 
\begin{pmatrix}
-A+C                & -\frac{B}{\sqrt{3}} & 0 & 0 \\
-\frac{B}{\sqrt{3}} & \frac{1}{3}(A-2B)+C & 0 & 0 \\
0 & 0 & -\frac{1}{3}(A + 2B)+C  & -\frac{B}{\sqrt{3}} \\
0 & 0 & -\frac{B}{\sqrt{3}}     & A+C
\end{pmatrix}
v_\Delta \,, \notag \\
&= v_{\bar{\Delta}}\bigg\{C\,\mathbb{1}_{4\times4} + K_\Delta\bigg\}
   v_\Delta \,,
\end{align}
where the vectors $v_{\bar{\Delta}}$ and $v_\Delta$ are vectors of the
(QCD) delta basis states, 
\begin{equation}\label{E:QCDeigenV}
v_{\bar{\Delta}} = 
\begin{pmatrix}
\bar{\D}^{++} & \bar{\D}^0 & \bar{\D}^+ & \bar{\D}^-
\end{pmatrix} \,, \qquad\qquad
v_\D = 
\begin{pmatrix}
\D^{++} & \D^0 & \D^+ & \D^-
\end{pmatrix}\transpose \,, 
\end{equation}
and
\begin{equation}
K_\Delta = 
\begin{pmatrix}
-A & -\frac{B}{\sqrt{3}} & 0 & 0 \\
-\frac{B}{\sqrt{3}} & \frac{1}{3}(A-2B) & 0 & 0 \\
0 & 0 & -\frac{1}{3}(A + 2B) & -\frac{B}{\sqrt{3}} \\
0 & 0 &  -\frac{B}{\sqrt{3}} & A
\end{pmatrix} \,.
\end{equation}
The entries in $M_\D$ are given by
\begin{align}
A &= 2\,\e_q\, \left( \g_M + t_1^{WM_+} 
      \frac{a \L_{QCD}^2}{\L_\chi}\cos(\w) \right) \,,
\qquad\qquad
B = t_2^{W_-}a^2\L_{QCD}^3\sin^2(\w) \,, 
\notag\\
C &= 2\,m_q\left(\g_M - 2\,\ol{\s}_M\right)
    -4\,\ol{\s}_W a\,\L_{QCD}^2\cos(\w)
    +2\,m_q\left(t_1^{WM_+} + 2\,t_2^{WM_+}\right)
     \frac{a \L_{\rm QCD}^2}{\L_\chi}\cos(\w) 
\notag\\
&\quad
+a^2\L_{QCD}^3\,(t +t_v) 
+a^2\L_{QCD}^3\left( 4\,t_1^{W_+}\cos^2(\w)
                    -2\,t_1^{W_-}\sin^2(\w)\right) \,. 
\end{align}
Note that to the accuracy we work, $\w$ can be either $\w_0$ or the
non-perturbatively determined twist angle. 

Except for the operators
\begin{equation*}
(\ol{T}_\mu \c{M}_+ T_\mu) \,,\qquad 
(\ol{T}_\mu\c{M}_+ T_\mu)\,{\rm tr}(\c{W}^{tw}_+) \,,\qquad
\ol{T}^{kji}_\mu(\c{W}^{tw}_-)^{ii'}(\c{W}^{tw}_-)^{jj'}T^{i'j'k}_\mu
\,, 
\end{equation*}
which contribute to $K_\Delta$, all other operators listed in
(\ref{E:Mtree}) above have trivial flavor structure, and so contribute
to the identity part of $M_\Delta$. Hence, to diagonalize $M_\Delta$,
we need only diagonalize $K_\Delta$. The orthogonal matrix that
accomplishes this is
\begin{equation} \label{E:Smtx}
S = 
\begin{pmatrix}
\frac{(2A-B+2X_-)^{1/2}}{2X_-^{1/2}} &
-\frac{(-2A+B+2X_-)^{1/2}}{2X_-^{1/2}} & 0 & 0 \\
\frac{\sqrt{3}B}{2X_-^{1/2}(2A-B+2X_-)^{1/2}} &
\frac{\sqrt{3}B}{2X_-^{1/2}(-2A+B+2X_-)^{1/2}} & 0 & 0 \\
0 & 0 & \frac{(2A+B+2X_+)^{1/2}}{2X_+^{1/2}} & 
       -\frac{(-2A-B+2X_+)^{1/2}}{2X_+^{1/2}} \\
0 & 0 & \frac{\sqrt{3}B}{2X_+^{1/2}(2A+B+2X_+)^{-1/2}} &
        \frac{\sqrt{3}B}{2X_+^{1/2}(-2A-B+2X_+)^{-1/2}} 
\end{pmatrix} \,,
\end{equation}
where $X_\pm = \sqrt{A^2 \pm AB + B^2}$, and each column of $S$ is a 
normalized eigenvector of $M_\Delta$ (and hence $K_\Delta$ also). The
diagonal matrix one obtains after diagonalizing $M_\Delta$ is then
\begin{equation} \label{E:Dmtx}
\c{D} = S^{-1} \cdot M_\D \cdot S = 
\frac{1}{3}\,{\rm diag}
\begin{pmatrix}
-A - B - 2X_- + 3C \\ 
-A - B + 2X_- + 3C \\ 
\!\quad A - B - 2X_+ + 3C \\ 
\!\quad A - B + 2X_+ + 3C 
\end{pmatrix} \,,
\end{equation}
where each entry in $\c{D}$ is an eigenvalue of $M_\Delta$.

Now if $\epsilon_q \neq 0$, $A \neq 0$. Hence, since in our power
counting $A \sim \mathcal{O}(\varepsilon^2)$ and 
$B \sim \mathcal{O}(\varepsilon^4)$, we may expand $X_\pm$ in the
ratio of $B/A \sim \mathcal{O}(\varepsilon^2) \ll 1$ as
\begin{equation}
X_\pm = A\sqrt{1 \pm \frac{B}{A} + \frac{B^2}{A^2}} 
      = A\left(1 \pm \frac{1}{2}\frac{B}{A} + \frac{3}{8}\frac{B^2}{A^2}
        +\c{O}(\varepsilon^6)\right) \,,
\end{equation}
from which it follows that
\begin{equation} \label{E:Spert}
S = 
\begin{pmatrix}
1 & -\frac{\sqrt{3}B}{4A} & 0 & 0 \\
\frac{\sqrt{3}B}{4A} & 1+\frac{B}{4A} & 0 & 0 \\
0 & 0 & 1 & -\frac{\sqrt{3}B}{4A} \\
0 & 0 & \frac{\sqrt{3}B}{4A} & 1-\frac{B}{4A}
\end{pmatrix} \,, \qquad\qquad
\mathcal{D} = {\rm diag}
\begin{pmatrix}
-A + C \\
\!\quad\frac{A}{3} - \frac{2B}{3} + C \\
-\frac{A}{3} - \frac{2B}{3} + C \\
\!\quad A + C \\
\end{pmatrix} \,, 
\end{equation}
up to corrections of $\mathcal{O}(\varepsilon^4)$ for $S$ and
$\mathcal{O}(\varepsilon^6)$ for $\c{D}$.  

If $\e_q = 0$, i.e. in the isospin limit, $A = 0$ and 
$X_\pm = B$. In this case, one can not find $S$ and $\c{D}$ in the
isospin limit by taking the limit $A \rightarrow 0$ in~\eqref{E:Spert},
since expansion in the ratio of $B/A$ is clearly not valid. Instead,
one has to go back to Eq.~\eqref{E:Smtx} and Eq.~\eqref{E:Dmtx}, which
in the isospin limit reduce to   
\begin{equation} \label{E:zeroA}
S = 
\begin{pmatrix}
\frac{1}{2} & -\frac{\sqrt{3}}{2} & 0 & 0 \\
\frac{\sqrt{3}}{2} & \frac{1}{2} & 0 & 0 \\
0 & 0 & \frac{\sqrt{3}}{2} & -\frac{1}{2} \\
0 & 0 & \frac{1}{2} & \frac{\sqrt{3}}{2} 
\end{pmatrix} \,, \qquad\qquad
\mathcal{D} = {\rm diag}
\begin{pmatrix}
-B + C \\ 
\;\;\,\frac{B}{3} + C \\ 
-B + C \\ 
\;\;\,\frac{B}{3} + C
\end{pmatrix} \,,
\end{equation}
and the eigenvalues contained in $\c{D}$ given in Eq.~\eqref{E:zeroA}
are the masses of the deltas at tree level in the isospin limit given in
Eq.~\eqref{eq:DLO} and Eq.~\eqref{eq:DNNLOtau1}. Note that as
discussed in the text, in the isospin limit, we need not perform any
mass matrix diagonalization at all, since we can simply rotate from
the outset to the basis where the twist is implemented by the diagonal
$\tau_3$.  

The new delta basis states are defined by 
\begin{equation}
v'_\Delta = S^{-1} \cdot v_\Delta \,,\qquad 
v'_{\bar{\Delta}} = v_{\bar{\Delta}} \cdot S \,,
\end{equation} 
in which the delta mass matrix is diagonal to the order we work.
By writing the old (unprimed) delta basis states in terms of the new
(primed) basis states using the defining relations given above, i.e.
\begin{equation}
v_\Delta = S \cdot v'_\Delta \,,\qquad 
v_{\bar{\Delta}} = v'_{\bar{\Delta}} \cdot S^{-1} \,,
\end{equation} 
the Lagrangian in the new delta basis can be obtained. Note that in
the case where $\epsilon_q \neq 0$, the new basis states contained in $v'_\Delta$ are ``perturbatively close'' to those contained in the $v_\Delta$, i.e. the
difference is $\c{O}(\varepsilon^2)$ as can be easily seen from
Eq.~\eqref{E:Spert}. This is of course not true if we are in a region
where $B \sim A$, or $\epsilon_q \sim a^2\L_{\rm QCD}^3$.

\end{document}

%% file: def.tex
\def\a{{\alpha}}
\def\b{{\beta}}
\def\d{{\delta}}
\def\D{{\Delta}}
\def\e{{\epsilon}}
\def\g{{\gamma}}
\def\G{{\Gamma}}
\def\k{{\kappa}}
\def\l{{\lambda}}
\def\L{{\Lambda}}
\def\m{{\mu}}
\def\n{{\nu}}
\def\w{{\omega}}
\def\O{{\Omega}}
\def\S{{\Sigma}}
\def\s{{\sigma}}
\def\t{{\tau}}
\def\th{{\theta}}
\def\x{{\xi}}

\def\ol#1{{\overline{#1}}}

\def\Dslash{D\hskip-0.65em /}
\def\dslash{{\partial\hskip-0.5em /}}
\def\vslash{{\rlap \slash v}}
\def\qbar{{\overline q}}

\def\CPT{{$\chi$PT}}
\def\QCPT{{Q$\chi$PT}}
\def\PQCPT{{PQ$\chi$PT}}
\def\tr{\text{tr}}
\def\str{\text{str}}
\def\diag{\text{diag}}
\def\order{{\mathcal O}}
\def\vit{{\it v}}
\def\vD{\vit\cdot D}
\def\am{\alpha_M}
\def\bm{\beta_M}
\def\gm{\gamma_M}
\def\smb{\sigma_M}
\def\smt{\overline{\sigma}_M}
\def\tb{{\tilde b}}

\def\c#1{{\mathcal #1}}

\def\Bbar{\overline{B}}
\def\Tbar{\overline{T}}
\def\cBbar{\overline{\cal B}}
\def\cTbar{\overline{\cal T}}
\def\pq{(PQ)}

\def\eqref#1{{(\ref{#1})}}